\documentclass[aps,pra,numerical,endfloats,preprint]{revtex4-1}
\usepackage{graphicx}
\usepackage{amsmath}
\usepackage{setspace}
\usepackage{amsfonts}
\usepackage[dvips]{epsfig}

\begin{document}
\title{Feshbach resonances and weakly bound molecular states of boson-boson and boson-fermion NaK pairs}
\author{Alexandra Viel}
\email{alexandra.viel@univ-rennes1.fr}
\author{Andrea Simoni}
\email{andrea.simoni@univ-rennes1.fr}
\affiliation{Institut de Physique de Rennes, UMR 6251, CNRS \& Universit\'e de Rennes 1, F-35042 Rennes, France}
\date{\today}

\begin{abstract} 
We study theoretically magnetically induced Feshbach resonances and
near-threshold bound states in isotopic NaK pairs. Our calculations
accurately reproduce Feshbach spectroscopy data on Na$^{40}$K and explain the origin of
the observed multiplets in the $p$-wave [Phys. Rev. A {\bf 85}, 051602(R) (2012)]. We apply the model to
predict scattering and bound state threshold properties of the boson-boson
Na$^{39}$K and Na$^{41}$K systems. We find that the Na$^{39}$K isotopic pair
presents broad magnetic Feshbach resonances and favorable ground-state
features for producing non-reactive polar molecules by two-photon
association. Broad $s$-wave resonances 
are also predicted for Na$^{41}$K collisions.
\end{abstract}
\maketitle
\section{Introduction}\label{sec_intro}

Ultracold gases are extraordinary systems to investigate fundamental
quantum phenomena in a highly controllable environment leading to a
wealth of spectacular experimental and theoretical results. More than
a decade ago the experimental production of ultracold mixtures of
alkali gases added a new twist to the cold atom field, paving the
way towards the study of few- and many-body phenomena absent in a pure
homonuclear gas. Few examples include recent experiments with polaronic
impurities~\cite{2009-AS-PRL-230402,2012-JC-PRA-023623,2014-CK-NAT-615},
formation of chemically reactive~\cite{2008-KKN-SCI-322}
or non-reactive~\cite{2014-TT-PRL-205301} ultracold
polar molecules, theoretical studies of phase
diagrams~\cite{2006-KM-PRL-120407} and of pairing in imbalanced Fermi
systems~\cite{2008-GGB-PRL-116405,2010-GO-PRL-065301}.

In this context, magnetic Feshbach resonances (FR) proved to
be a powerful and versatile tool to widely tune few-body
interactions~\cite{2010-CC-RMP-1225}, allowing one to explore in a
controlled way regimes from the non interacting ideal behavior to strongly
interacting systems. A FR also offers the possibility
to associate pairs of atoms in  weakly bound molecular states using
time-dependent magnetic fields~\cite{2004-KG-PRB-3457}. Such molecules
have an intrinsic interest due to their long-range nature. Moreover,
depending on their spin and spatial structure they can be used as a
convenient initial state for producing polar molecules in the ground state
via stimulated Raman processes~\cite{2008-KKN-SCI-322,2014-TT-PRL-205301}.

Applications based on resonances require an accurate characterization
of the scattering dynamics and of the properties of bound states near
the dissociation threshold. Fortunately, theory can predict from a small
amount of experimental data the location and the width of magnetic
resonances and the relevant molecular state properties. In fact, to date
highly quantitative models exist for most alkali isotopic
pairs~\cite{2010-CC-RMP-1225}. In this work we focus on NaK mixtures,
a system composed of two species that can be individually cooled to
ultracold temperatures.

Experiments with Na and the fermionic isotope $^{40}$K are
currently being performed at MIT, where heteronuclear FR
spectra have been discovered and interpreted based on simplified
asymptotic models~\cite{2012-JWP-PRA_051602}. Magnetic 
association~\cite{2012-CHW-PRL-085301} and, more recently, transfer to
the ro-vibrational ground state of the dimer have also been
demonstrated~\cite{2015-JWP-PRL-205302}. Accurate Born-Oppenheimer
potentials for the ground and the excited states have been built and
used to study the adiabatic transfer of a Feshbach molecule to the
ground ro-vibrational state~\cite{2013-TAS-PRA-023401}. However, a
comprehensive account of scattering and bound state features for this
boson-fermion mixture in the electronic ground state is still lacking. In
addition, near-threshold properties for the boson-boson pairs Na$^{39}$K
and Na$^{41}$K are still unknown. This work aims therefore on one side
at giving a more complete picture of the Feshbach physics of Na$^{40}$K,
and on the other at providing theoretical predictions for the two purely
bosonic pairs, for which experiments are on the way in few groups worldwide. We
study both scattering and bound states for an extensive set of hyperfine
states, and discuss the experimental implications of our results for
interaction control and molecule production.

The paper is organized as follows. Sec.~\ref{sec_theory} introduces
our theoretical approach and the Born-Oppenheimer potentials optimization
procedure based on known experimental data. Sec.~\ref{sec_results}
presents results and discussions for the boson-fermion and the
boson-boson pairs. Few experimental implications of our
results are discussed. A conclusion \ref{sec_conclusion} ends this work.

\section{Methods}
\label{sec_theory}

\subsection{Computational approach}

We solve the time-independent Schr{\"o}dinger equation for bound and
scattering states in the well-known framework of the close-coupling
approach to molecular dynamics. Briefly, in our approach a basis of
Hund's case (b) molecular states is used to expand the
total wavefunction at each value of the interatomic distance $R$. In
Hund's case (b) the spin state of the dimer is represented as $| S M_S
I M_I \rangle$ with $\vec S$ and $\vec I$ the electronic and nuclear
spin angular momentum, respectively~\cite{lefebvrebrion2004}. The description of the diatomic is
completed by assigning the $\ell$ and $m$ quantum numbers relative to
the orbital angular momentum $\vec \ell$ of the atoms about their center
of mass. In this basis the electrostatic Born-Oppenheimer potentials are
represented by diagonal matrices with entries the singlet $X^1 \Sigma^+$
and triplet $a^3 \Sigma^+$ molecular symmetry adiabatic potential energy
curves (see below). 

The other interactions relevant for the ultracold regime included in this
work are the atomic hyperfine interaction $\mbox{$H_{\rm hf}=\sum_{k=a,b}
A_k {\vec s}_k \cdot {\vec \imath}_k $}$, the anisotropic spin-spin coupling
$\mbox{$H_{\rm ss}=\alpha^2  R^{-3} \left[ { \vec s}_a \cdot { \vec s}_b - 3
({ \vec s}_a \cdot {\hat R}) ( { \vec s}_b \cdot {\hat R} ) \right] $}$,
and the Zeeman interaction energy with the external magnetic field $H_{\rm
Z}=\mu_B \sum_{k=a,b} (g_s {\vec s}_{k} + g_i {\vec \imath}_{k}) \cdot {\vec B}
$. Here ${\vec s}_k$ and ${\vec \imath}_k$ are the electron and nuclear spin
of the individual atoms and $A_k$ the respective ground-state atomic hyperfine
constant for atoms $a$ and $b$, $g_s$ and $g_i$ the electron and nuclear gyromagnetic ratios,
$\alpha$ and $\mu_B$ the fine structure constant and the Bohr magneton
respectively. Such atomic interactions can be expressed in the molecular Hund's case (b)
computational basis using standard methods of angular momentum algebra
(see {\it e.g.}~\cite{1996-ET-JRNIST-505}).

Bound state calculations are performed using a variable grid approach
allowing one to represent over a sufficiently small number of points
the rapid oscillations at short range and the long range tail of the
dimer wave function~\cite{1998-ET-PRA-4257}. 
For scattering calculations we use the variable-step Gordon propagation or
the renormalized Numerov algorithm to efficiently solve the coupled-channel
Schr{\"o}dinger equation~\cite{1973-FHM-PRA-957}. Once the solution has
been built in the computational basis, a frame transformation is applied
to express the solution in an asymptotically diagonal representation
before using a standard matching procedure to extract the reactance
matrix $K$~\cite{1973-FHM-PRA-957}.

\subsection{Optimization of the molecular potential}
\label{sec_potential}

We adopt for this work the $X^1\Sigma^+$ and the $a^3\Sigma^+$
electronic ground state potential of the NaK molecule proposed in
Ref.~\cite{gerdes2008}. A minor modification is made to ensure a
continuous and continuously differentiable expression by fine tuning
the parameters given in~\cite{gerdes2008}. First, starting from the
asymptotic long range $U_{LR}(R)$ expressions we numerically enforce 
continuity at the switching points $R_i$ and $R_0$. In addition, smooth
damping functions are preferred to the published abrupt change of the
potential between the short range repulsive part, the inner well and the long range
part.  The resulting continuously differentiable expression is given by
\begin{eqnarray}
\label{eq_PES}
V(R)& = &U_{SR}(R)~\left[1-\tanh (\beta(R-R_i))\right]~\left[1-\tanh (\beta(R-R_0))\right]/4  \nonumber \\ 
 & & + U(R)~\left[1+\tanh (\beta(R-R_i))\right]~\left[1-\tanh (\beta(R-R_0))\right]/4  \nonumber \\
 & & + U_{LR}(R)~\left[1+\tanh (\beta(R-R_i))\right]~\left[1+\tanh (\beta(R-R_0))\right]/4,
\end{eqnarray}
where the parameterized functions $U_{SR}$, $U(R)$ and $U_{LR}(R)$ and
the switching points $R_i$ and $R_0$ are taken from the work of Tiemann
and coworkers \cite{gerdes2008}. Note there is a typo in the Tab.~1
of Ref.~\cite{gerdes2008} where $10^6$ should be replaced by $10^8$
for the $B$ constant value. A value of $80~a_0^{-1}$ has been found to be
suitable for the control parameter $\beta$ of the damping function
for the two electronic states. With an infinite $\beta$ value
the original potential curves are recovered.

We are now in the position to perform close-coupling calculations for
different initial channels. We will conventionally label each asymptotic
channel by specifying the separated-atom NaK state $|f_a,m_{f_a}\rangle
+ |f_b,m_{fb}\rangle$ with ${\vec f}_k={\vec s}_k + {\vec \imath}_k$
($k=a,b$) to which the latter adiabatically correlate as $B$ tends
to zero.
FR have been experimentally observed in the collision between Na in
$|1,1\rangle$ and $^{40}$K in $|9/2,m_f\rangle$, with $m_{f_b}=-9/2,
-7/2, -5/2$ and $-3/2$~\cite{2012-JWP-PRA_051602}. We compute the $s$-wave
scattering length $a$ in the relevant channels and search for resonances
as poles of $a$ as a function of magnetic field.
We also search for $p$-wave resonances by locating maxima in the partial
$p$-wave elastic cross sections at a fixed collision energy of $1~\mu$K.
The observed resonance locations are not reproduced accurately by the
original potentials. However, a simple modification consisting in
introducing the correction terms
\begin{equation}
\label{eq_Vcorr}
V_{corr}^{(i)}(R)=s^{(i)}/\cosh\left[\frac{R-R_{e}^{(i)}}{R_{0}^{(i)}}\right]^2, \mbox{ for } i=1,3,
\end{equation}
near the bottom of the $X^1\Sigma^+$ and $a^3\Sigma^+$ electronic
potentials enables us to model the experimentally measured
spectra. More specifically, using $R_{e}^{(1)}=6.61$~$a_0$,
$R_{e}^{(2)}=10.29$~$a_0$ and $R_{0}^{(2)}=1.5$~$a_0$ a
Levenberg-Marquardt algorithm is applied to determine two optimal
$s^{(i)}$ parameters 
\begin{subequations}
\begin{eqnarray}
\label{eq_shift1_shift3}
s^{(1)} &  = & -0.37247~10^{-4} ~ E_h \\
s^{(3)} &  = & -0.16385~10^{-5} ~ E_h .
\end{eqnarray}
\end{subequations}
In the fitting procedure, we included three $s-$wave resonances at 78.3~G,
88.2~G and 81.6~G and an average position of the $p$-wave resonances
appearing as a multiplet around 19.19~G \cite{2012-JWP-PRA_051602}. For
the $p$-wave calculation we initially neglect the spin-coupling term
responsible of the multiplet structure thus avoiding possible incorrect
labeling of the closely spaced $p$-wave resonances.

After optimization, the resonances positions are theoretically reproduced
with a reduced $\chi^2$ = 0.57. On a more physical ground, the artificial
control parameters are usually translated in corresponding singlet and
triplet scattering length $a_{\rm S,T}$ (see Tab.~\ref{tab_ast}). Our
optimized boson-fermion model can also be used for predicting the
properties of boson-boson isotopes. Within the Born-Oppenheimer
approximation, which is expected to be accurate for all but for the
lightest species~\cite{2014-PSJ-PRA-052715}, it is sufficient to change
the value of the reduced mass in the Hamiltonian to compute $a_{\rm
S,T}$. Note however that if the number of bound states in our nominal
potentials turned out to be incorrect, the predicted $a_{\rm S,T}$
(and hence the results of the coupled model) will be
systematically shifted.
\begin{table}[hb!t]
\caption[]{Singlet and triplet scattering lengths $a_{\rm S,T}$ obtained according to our optimized potentials for different NaK isotopic pairs.}
\label{tab_ast}
\begin{ruledtabular}
\begin{tabular}{ccc}
Isotope  & $a_{\rm S}(a_0)$ &  $a_{\rm T}(a_0)$    \\ \hline
Na$^{39}$K  & 255     &    -84              \\
Na$^{40}$K  & 63      &   -838            \\
Na$^{41}$K  & -3.65  &    267
\end{tabular}
\end{ruledtabular}
\end{table}

We begin our analysis with a detailed discussion of the resonances for
the Na$^{40}$K boson-fermion mixture in the section~\ref{sec_nak40} here after.
\begin{table}[!bth]
\caption[]{List of the $s$-wave FR for Na$^{40}$K. Our theoretical
resonance positions $B_\text{res}$ are compared with the experimental
data $B_\text{exp}$ from Ref.~\cite{2012-JWP-PRA_051602}. The locations
where the scattering length vanishes $B_\text{ZC}$, the effective range on resonance $r_\text{eff}^\text{res}$,
the resonance strength parameter $s_\text{res}$ and the fitting parameters $a_{bg}$, $\varepsilon$, and $\Delta$ defined in
Eq.~\eqref{eq_a_width} are reported in columns 4-9. See text for more details.}
\label{tab_nak40}
\begin{ruledtabular}
\begin{tabular}{ccrrclrrr}
Na$^{40}$K channel          & $B_\text{exp}~(G)$             & $B_\text{res}~(G)$ & $B_\text{ZC}~(G)$ & $r_\text{eff}^\text{res}~(a_0)$ & $s_\text{res}$  & $a_{bg}~(a_0)$ & $\varepsilon ~(G^{-1})$& $\Delta~(G)$  \\ \hline
$|1,1\rangle$ +$|9/2,-9/2\rangle$  &78.3 &  77.78  & 72.23 &    -0.668        &   0.682    & -619.3 &-1.2~10$^{-3}$ & -5.5   \\
                                   &88.2 &  88.68  & 79.82 &   137.           &   9.92     &        &                &-8.8   \\ \hline
$|1,1\rangle$ +$|9/2,-7/2\rangle$  &81.6 &  81.42  & 81.18 & -8.39~10$^4$     &   0.0120   & -552.7 &-1.1~10$^{-3}$ &-0.23   \\
                                   &89.8 &  89.82  & 83.61 &   -48.4          &   0.517    &        &                &-6.2   \\
                                   &108.6& 108.91  & 96.86 &   141.           &  12.3      &        &                &-16.0   \\ \hline
$|1,1\rangle$ +$|9/2,-5/2\rangle$  &96.5 &  96.39  & 95.75 & -4.84~10$^4$     &   0.0205   & -496.4 &-1.0~10$^{-3}$  & -0.6   \\
                                   &106.9& 106.54  & 98.83 &   -90.9          &   0.426    &        &                & -7.5   \\
                                   &138  & 136.82  &110.53 &   142.           &  14.58     &        &                & -26.2   \\ \hline
$|1,1\rangle$ +$|9/2,-3/2\rangle$  &116.9& 117.19  &115.62 & -3.46~10$^4$     &   0.0283   & -443.1 &-8.5~10$^{-4}$ &-1.2  \\
                                   &129.5& 130.36  &119.85 &  -120.           &   0.379    &        &                &-9.8  \\
                                   &175  & 177.44  &135.35 &   143.           &  17.3      &        &                &-41.7   \\
\end{tabular}
\end{ruledtabular}
\end{table}

\section{Results} \label{sec_results}
\subsection{Na$^{40}$K} \label{sec_nak40}
We now perform extensive close-coupling calculations with the optimized potentials
described in Sec.~\ref{sec_potential}. Table~\ref{tab_nak40}
summarizes our findings for the $s$-wave for magnetic fields up to
1000~G for different hyperfine levels. 
We report in the table the positions of the poles $B_\text{res}$
observed in the calculated $s$-wave scattering length $a$ as well as the nearby
zero-crossing field $B_\text{ZC}$ where $a$ vanishes. The experimental
data of Ref.~\cite{2012-JWP-PRA_051602} are also reproduced in the table.

The good quality of the theoretical model after optimization is confirmed
by the very good agreement (below $0.5$~G) illustrated in the table
for all narrow features experimentally observed in different hyperfine
combinations. A larger discrepancy of $\sim 1$G is found on broader
resonances which may however be more difficult to locate experimentally
with accuracy. No additional $s$-wave features are found with respect
to the experiment.

In order to extract the magnetic width $\Delta$, the scattering length $a(B)$ obtained for each incoming channel is
fitted according to a formula appropriate for overlapping resonances \cite{jachymski2013},
\begin{equation}
\label{eq_a_width}
a(B)=a_{bg}(1+\varepsilon B) \prod_i^N\left(1-\frac{\Delta_i}{B-B_{{\rm res},i}}\right),
\end{equation}
in which a linear variation of $a_{bg}$ as a function of the magnetic
field is assumed. 
Eq.~\eqref{eq_a_width} reduces to the well-known standard
expression $a(B)=a_{bg}\left[ 1-\Delta/(B-B_\text{res}) \right]$ if resonances are
isolated and the background scattering length is locally constant. The
fitting of the scattering length with Eq.~(\ref{eq_a_width}) has been
carried out for magnetic fields spanning a $\pm 4\Delta$~G region
around each resonance.  For overlapping cases, the largest $\Delta$
was taken to define the fitting interval.  We only report in the table
the corresponding $\Delta_i$, $a_{bg}$ and $\varepsilon$ parameters for
which Eq.~\eqref{eq_a_width} reproduces the numerical data to an accuracy
better than 5\% in either the relative error or in the absolute error
measured in units of van der Waals radius
\begin{equation}
R_\text{vdW}=\frac{1}{2}\left( \frac{2\mu C_6}{\hbar^2}\right)^{1/4}.
\end{equation}
The latter quantity ($\approx 53~a_0$ for NaK) represents the natural value of $a$
for scattering in a van der Waals potential~\cite{2010-CC-RMP-1225}.

In addition, for each resonance we extract the
effective range $r_\text{eff}$ defined through the low energy expansion of the
elastic reactance matrix element
\begin{equation}
\label{eq_reff}
\frac{k}{K(E,B)}= -\frac{1}{a(B)}+\frac{1}{2} r_\text{eff}(B) k^2
\end{equation}
computed {\it at the resonance value} $r_\text{eff}^\text{res}\equiv r_\text{eff}(B_\text{res})$ by a linear
fit of $k/K(E,B_\text{res})$ as a function of collision energy $E= \hbar^2 k^2 /(2 \mu)$ in an appropriate energy
domain. We introduce a corresponding intrinsic resonance length $R^*$
to characterize the resonance strength defined as 
\begin{equation}
\label{eq_reso_length}
r_\text{eff}=-2 R^* +\frac{2}{3 \pi} \Gamma^2 \left( \frac{1}{4} \right) R_\text{vdW}  .
\end{equation}
In the case of isolated resonances the resonance length can be expressed in terms of
scattering background and resonance parameters and of the magnetic moment difference $\delta \mu$ between the
open and the closed channel as $R^*=\hbar^2 / (2 \mu a_{\rm bg} \delta \mu
\Delta)$~\cite{2010-CC-RMP-1225}. As remarked in the supplemental
material of Ref.~\cite{PhysRevLett.111.053202} the Eq.~\eqref{eq_reso_length}
also holds for overlapping resonances to the extent that they are not
directly interacting. According to the relative value of the length $R^*$
being much larger (resp. much smaller) than the van der Waals length,
resonances are classified as being open channel (resp. closed channel)
dominated. The
resonance strength parameter defined as
\begin{equation}
s_\text{res} = \frac{0.956 R_\text{vdW}}{R^*}
\end{equation}
and listed in Tab.~\ref{tab_nak40}
is therefore a useful dimensionless indicator of the resonance character~\cite{2010-CC-RMP-1225}.

It is interesting to compare the present close-coupling data with the
results of an asymptotic bound state model used for interpretation in
Ref.~\cite{2012-JWP-PRA_051602} (not shown in the table). Considering the
simplicity of the latter, the agreement is good as far as the resonance
position and the width of the largest features is concerned. The most
serious discrepancy bears on the width of the narrow $s$-wave resonances,
which are underestimated by more than one order of magnitude by the
asymptotic model. 

Feshbach molecules formed by magnetic association can be a good starting
point to form molecules in the absolute ro-vibrational ground state using
two-photon transfer schemes. A first constraint to be taken into account
to achieve such a transfer is that according to electric dipole selection
rules only Feshbach molecular states of significant singlet character
can be coupled to ground-state singlet molecules if singlet excited
states are used as a bridge. If the initial Feshbach molecule turns out
to have mostly triplet character one can use excited electronic states
of mixed singlet-triplet character as a bridge, an approach suggested
for NaK in~\cite{2013-TAS-PRA-023401} and successfully recently adopted
to form Na$^{40}$K in the absolute ground state by a STIRAP two-photon
process~\cite{2015-JWP-NJP-075016,2015-JWP-PRL-205302}. Moreover,
the radial overlap between the excited intermediate state and both the
initial and the target ground state molecule must be significant. To
gain more insight into the resonance nature and to get a hint at the
expected efficiency of two-photon processes we perform bound state
coupled-channel calculations. A detailed analysis having already been
performed in~\cite{2013-TAS-PRA-023401} for Na$^{40}$K, here we just
stress the main elements for the sake of comparison with the following
analysis of the boson-boson pairs.

\begin{figure}[!hbt]
\centerline{\epsfig{file=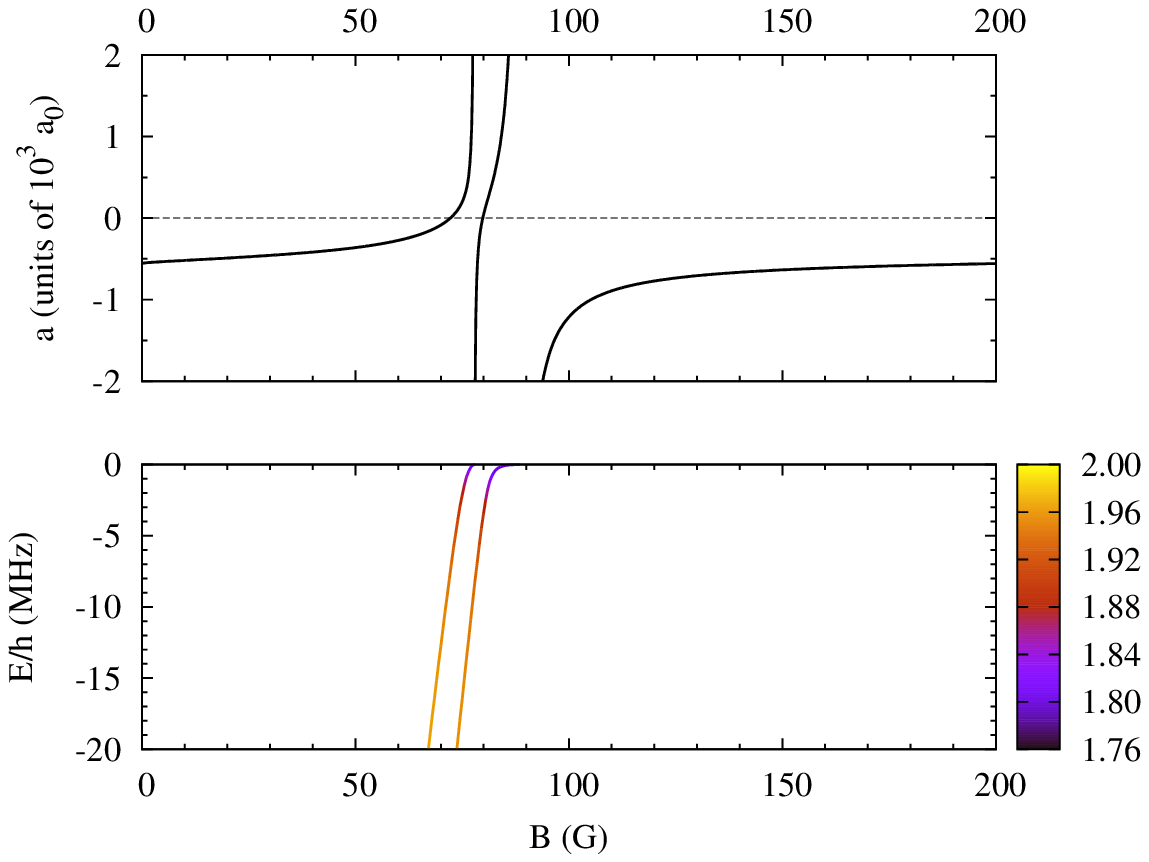,width=.99\columnwidth,angle=0}}
\caption[]{(color online) Scattering length $a$ as a function of the magnetic field $B$
for Na$|1,1\rangle$ + $^{40}$K$|9/2,-9/2\rangle$ $s$-wave collisions (top panel).
Corresponding molecular energy levels as a function
of $B$ are shown in the lower panel. 
The density code denotes the average spin $\langle {\vec S}^2 \rangle$ of the molecule. }
\label{fig_nak40_aa}
\end{figure}

We depict for instance in Fig.~\ref{fig_nak40_aa} the scattering length
and the evolution of the molecular levels near the broadest resonance in
the hyperfine absolute ground state which has been successfully used as
starting point for STIRAP association~\cite{2015-JWP-PRL-205302}. The
corresponding average electronic spin $\langle {\vec S}^2 \rangle$
is also shown as a function of internuclear distance as color
code in the lower panel. The nearly pure triplet character of this
molecular state is in principle unfavorable for the production of ground state
singlet molecules through the excited singlet manifold. Inspection of
lower panel of Fig.~\ref{fig_nak40_aa} might suggest working closer
to resonance to increase the single character as the molecular
state mixes with the scattering continuum. However, as already noted
in Ref.~\cite{2013-TAS-PRA-023401} such admixture comes to the price of
a delocalization of the wavefunction at larger distances, and thus to a
decreased overlap with the intermediate excited state. Such strong triplet
character is a common feature of the molecular states associated to all
broadest resonances in Tab.~\ref{tab_nak40}, which is in fact to a good
approximation a common molecular state in the triplet potential with a
different projection $m_f$ of the total hyperfine angular momentum $\vec
f = \vec S + \vec I$. In conclusion, use of bridge spin-orbit coupled
states to help enhance the transfer efficiency seems necessary for the
boson-fermion pair~\cite{2013-TAS-PRA-023401}. We will show below that 
the situation is significantly different
for the boson-boson mixtures.

\begin{figure}[!hbt]
\centerline{\epsfig{file=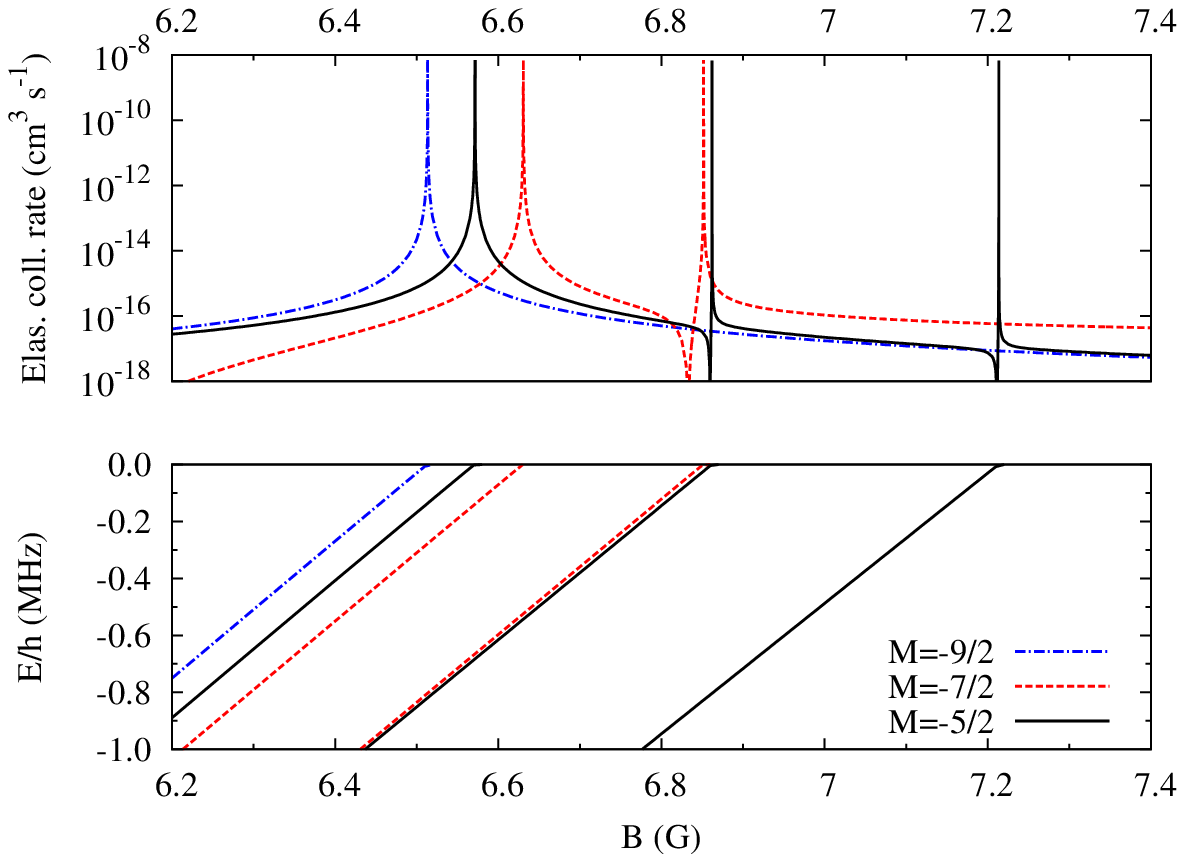,width=.99\columnwidth,angle=0}}
\caption[]{(color online) Elastic rate computed at a collision energy of $1~\mu$K as a function of the magnetic field $B$ for Na$|1,1\rangle$ + $^{40}$K$|9/2,-9/2\rangle$ $p$-wave collisions (top panel). Corresponding molecular energy levels as a function of $B$ are shown in the lower panel. Curves are resolved according to the projection $M$ of total angular momentum.} \label{fig_nak40_aa_p}
\end{figure}

In addition to the observed $s$- and $p$-wave features, additional
$p$-wave resonances are also predicted by our model. We compute the
elastic collision rate up to $B=1000$~G and present a restricted magnetic
field range in the upper panel of Fig.~\ref{fig_nak40_aa_p}. The lower panel
depicts the energy levels of the molecular states responsible for each resonance. 
Resonance features are detected by local maxima in the elastic collision rate as
well as in the inelastic probabilities. 
The position of these maxima agree to better than 0.02~G for all except the two features around 21.9~G in the $|1, 1\rangle + |9/2,-7/2\rangle$ channel for which the differences are 0.05~G. 
It turns out to be easier to extract the location of the resonances from the inelastic probabilities.
These positions are summarized in Tab.~\ref{tab_nak40_p} together with the position of the local maxima in the elastic collision rate when no inelastic process are present.

\begin{table}[hb!t]
\caption[]{Theoretical magnetic field locations $B_\text{res}$ of $p$-wave
FR for Na$^{40}$K for projection of total angular momentum 
$M$ in different hyperfine atomic channels. Calculations are performed at collision energy of $1~\mu$K. 
Available experimental values $B_\text{exp}$ are also reported \cite{2012-JWP-PRA_051602}.
}
\label{tab_nak40_p}
\begin{tabular}{p{8cm}ccp{8cm}}
\begin{ruledtabular}
\begin{tabular}{rrrr}
 Na$^{40}$K channel & $M$ & $B_\text{res}~(G)$ &$B_\text{exp}~(G)$  \\ \hline 
$|1, 1\rangle + |9/2,-9/2\rangle$ &  -9/2 &  6.51& 6.35  \\ \
                                  & -5/2 &  6.57& 6.41    \\
                                  & -7/2 &  6.63& 6.47    \\
                                  & -7/2 &  6.85& 6.68    \\
                                  & -5/2 &  6.86&         \\
                                  & -5/2 &  7.21&         \\
                                  & -7/2 & 18.01&         \\
                                  & -9/2 & 18.36&         \\
                                  & -5/2 & 19.32& 19.12   \\
                                  & -9/2 & 19.39& 19.18   \\
                                  & -7/2 & 19.47& 19.27   \\
                                  & -5/2 & 20.72&         \\
                                  & -7/2 & 20.82&         \\
                                  & -5/2 & 22.43&         \\ \hline 
$|1, 1\rangle + |9/2,-7/2\rangle$ & -5/2 &  7.55&         \\
                                  & -7/2 &  7.62&         \\
                                  & -7/2 &  7.88&  7.32   \\
                                  & -5/2 &  7.90&  7.54   \\
                                  & -3/2 &  7.90&         \\
                                  & -3/2 &  8.27&         \\
                                  & -5/2 &  8.33&         \\
                                  & -3/2 &  8.87&         \\
                                  & -7/2 & 20.09&         \\
                                  & -5/2 & 21.76&         \\
                                  & -7/2 & 21.91&         \\
                                  & -5/2 & 23.49& 23.19   \\
                                  & -3/2 & 23.53&         \\
                                  & -7/2 & 23.60& 23.29   \\
                                  & -3/2 & 25.33&         \\
                                  & -5/2 & 25.61&         \\
                                  & -3/2 & 28.01&         \\
\end{tabular}
\end{ruledtabular}
&~~~~~ & &
\begin{ruledtabular}
\begin{tabular}{rrrr}
Na$^{40}$K channel &  $M$ & $B_\text{res}~(G)$ &$B_\text{exp}~(G)$  \\ \hline 
$|1, 1\rangle + |9/2,-5/2\rangle$ &  -5/2 & 8.87  &        \\
                                  & -5/2 &  9.34 &  9.23   \\
                                  & -3/2 &  9.33 &         \\
                                  & -3/2 &  9.82 &  9.60   \\
                                  & -5/2 &  9.88 &         \\
                                  & -1/2 &  9.89 &         \\
                                  & -1/2 & 10.46 &         \\
                                  & -3/2 & 10.61 &         \\
                                  & -1/2 & 11.50 &         \\
                                  & -5/2 & 24.80 &         \\
                                  & -5/2 & 27.20 &         \\
                                  & -3/2 & 27.28 &         \\
                                  & -3/2 & 29.61 &  29.19  \\
                                  & -1/2 & 29.88 &  29.45  \\
                                  & -5/2 & 29.93 &  29.52  \\
                                  & -1/2 & 32.57 &         \\
                                  & -3/2 & 33.04 &         \\
                                  & -1/2 & 36.97 &         \\ \hline 
$|1, 1\rangle + |9/2,-3/2\rangle$ & -3/2 & 11.35&        \\
                                  & -1/2 & 11.95&        \\
                                  & -3/2 & 12.12&        \\
                                  & -1/2 & 13.04&12.51   \\
                                  & -3/2 & 13.20&        \\
                                  & 1/2  & 13.20&12.68   \\
                                  & 1/2  & 14.30&        \\
                                  & -1/2 & 14.55&        \\
                                  & 1/2  & 16.31&        \\
                                  & -3/2 & 32.00&        \\
                                  & -3/2 & 35.74&        \\
                                  & -1/2 & 36.10&        \\
                                  & -1/2 & 39.87&39.39   \\
                                  & 1/2  & 40.37&39.86   \\
                                  & -3/2 & 40.42&        \\
                                  &  1/2 & 45.18&        \\
                                  & -1/2 & 45.85&        \\
                                  &  1/2 & 53.21&        \\
\end{tabular}
\end{ruledtabular}
\end{tabular}
\end{table}
The $p$-wave multiplets observed in Ref.~\cite{2012-JWP-PRA_051602}
and reproduced in Tab.~\ref{tab_nak40_p} are at first sight
surprising since the spin-spin interaction typically gives rise to
doublets~\cite{PhysRevA.69.042712}. The nature and multiplicity of such
magnetic spectrum can be rationalized starting from a picture where the
spin interaction is at first neglected. In this situation the total internal spin projection $m_f=m_{f_a}+m_{f_b}$ is an
exactly conserved quantum number. Let us consider for definiteness the
case of two free atoms with $m_f= -7/2$. Let us moreover
restrict ourselves to $\ell =1$, since the $\ell >1 $ contributions
are vanishingly small at the present very low collision energies due to
centrifugal repulsion. The projection $M$ of the total angular momentum,
which is strictly conserved, can then only take values $-9/2$, $-7/2$,
and $-5/2$.  Within this restricted model and fixing $f=7/2$, one can
build six molecular states
 with projections $\{ m_f m \}= \{ -7/2 ; 0 ,\pm 1 \} ,\{ -5/2 ; 0,-1 \}
 , \{ -3/2 ; -1 \}$ that are
degenerate, since both $\vec f$ and $\vec \ell$ are strictly
conserved in the absence of anisotropic spin-spin and of the Zeeman
interaction~\cite{1996-ET-JRNIST-505}.

If $B=0$ and the spin-spin interaction does not vanish, conservation of
total angular momentum $\vec F= \vec f + \vec \ell$ guarantees that the
six molecular states will give rise to one
triply degenerate level with $F=9/2$ corresponding to $M=-9/2,-7/2,-5/2$, one doubly degenerate
level with $F=7/2$ and with $M=-7/2,-5/2$ and one singly degenerate level with $F=5/2$ and with $M=-5/2$.
The energy differences between the three groups is small due to the weakness of the spin-spin interaction.
However the mixing of the different $\{ m_f, m\}$ within each block of given $M$ can be strong.
\begin{figure}[!hbt]
\centerline{\epsfig{file=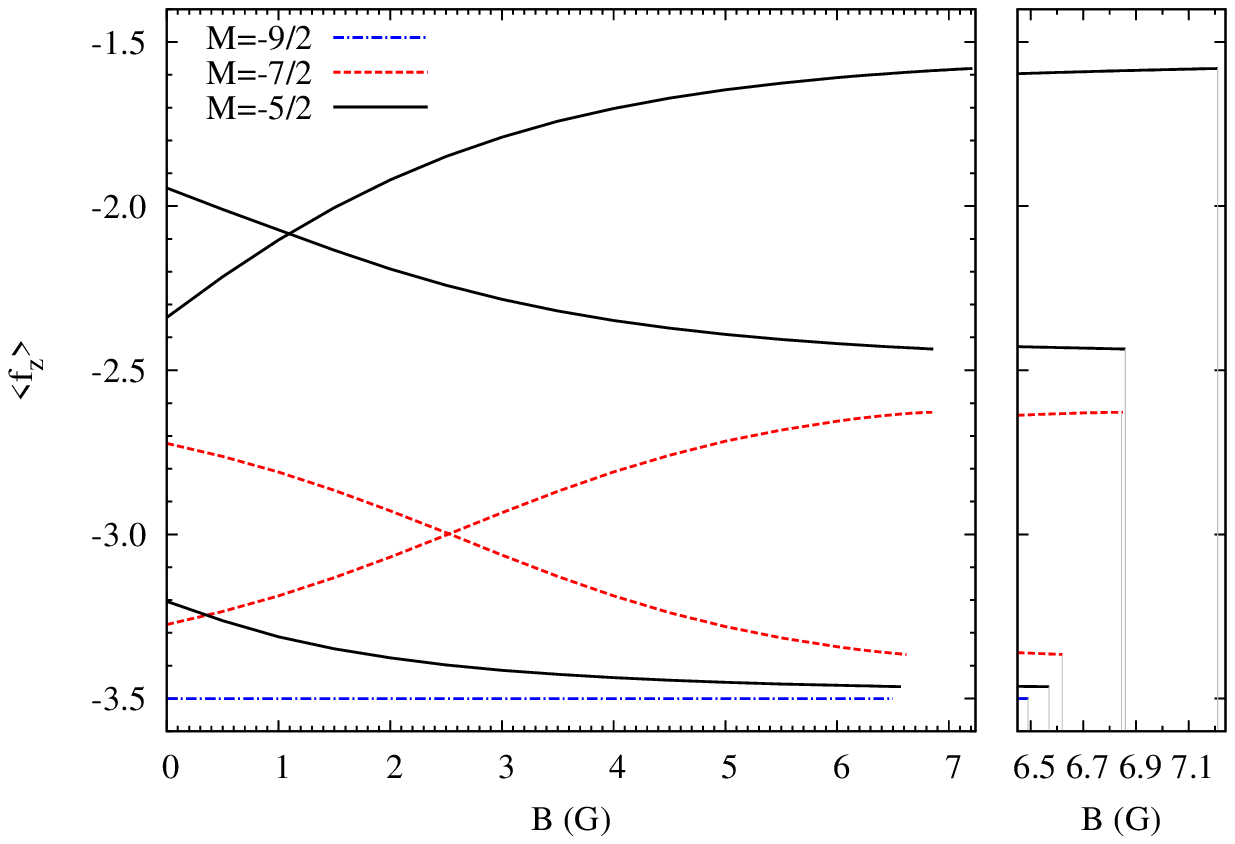,width=.99\columnwidth,angle=0}}
\caption[]{
(color online) Average value of the $z$ projection of the total internal
spin ${\vec f}={\vec f}_a+{\vec f}_b$ as a function of $B$ for $\ell=1$ bound states
with total angular momentum projection $M$. The right panel presents a
blowup for magnetic fields in the  6.4 to 7.2~G range.
}
\label{fig_nak40_aa_p_mf}
\end{figure}
In fact, as shown in Fig.~\ref{fig_nak40_aa_p_mf}, even in zero
field the values of $m_f$, and hence of $m$, is undefined with the
exception of the bound level with $M=-9/2$ which being essentially
isolated retains to high accuracy its $m_f=-7/2$ and $m=-1$ character.
Total rotational invariance and thus degeneracy with respect to $M$ is
broken by the magnetic field which introduces a coupling between states
with different quantum number $F$ in subspaces of given $M$. As the magnetic field
increases, the computed averaged projection $\langle f_z \rangle $ in
the $\vec B$ direction converges for each of the six molecular states 
toward the large $B$ limit, that is three levels with $m_f=-7/2$,
two with $m_f=-5/2$ and one with $m_f=-3/2$. 

This limit is however not fully reached when the bound state energies
cross threshold giving rise to resonance.  We conclude that for the
magnetic field values of relevance for the resonances the spin-spin
perturbation is comparable with the Zeeman splitting. A treatment
of the spin-spin interaction to first order perturbation theory in
subspaces spanned by states of given $m_f$ and $m=0,\pm1$ is thus not
appropriate in the present case.  As a consequence, one cannot reach
the usual conclusion that states with same $|m|$ are degenerate, such
approximate degeneracy being lifted by the perturbing effect of nearby
states with different $m_f$.

Generalizing the argument above to other hyperfine combinations, we expect
a multiplet of six resonances at low $B$ followed by a multiplet with
eight resonances for the $|1, 1\rangle + |9/2,-9/2\rangle$ channel. The
$|1, 1\rangle + |9/2,-7/2\rangle$ channel present an eight-fold multiplet
at lower $B$ and a nine-fold multiplet for larger $B$ values. 
The two remaining channels, $|1, 1\rangle + |9/2,-5/2\rangle$ and $|1, 1\rangle + |9/2,-3/2\rangle$ present two nine-fold multiplets each.
Some resonances are only weakly coupled and do not result in marked peaks in the elastic collision rate.
More precisely, two molecular state crossings do not give rise to detectable features in the numerical elastic rate for the $|1, 1\rangle + |9/2,-5/2\rangle$ channel (at 8.87~G and at 24.80~G).
For the $|1, 1\rangle + |9/2,-3/2\rangle$ channel, five molecular state crossings have no detectable effects on the numerical rate (at 11.35, 11.95, 32.00 35.74 and 36.10~G).
These features are however clearly seen in the inelastic probabilties.

We now propose an assignment of $p$-wave resonances in the MIT experiment.
First of all, one may notice that the experimental spectrum only entails a subset of the predicted multiplets. 
Some features observed in the theoretical model (for instance, the pair near 6.85~G or the one near 7.90~G) are strong but nearly overlapping, such that one can reasonably assume that they have not been resolved in the experiment. 
In such cases, for the assignment we only retain the strongest of the two features in the theoretical spectrum.
Next, we affect the strongest elastic theoretical features to the experimental positions under the condition that the resulting splitting agrees with the experimental one.
The procedure is succesfull in all cases, with the exception of the low $B$ spectrum in the $| 1,1 \rangle + | 9/2 -5/2 \rangle $ channel. 
Note however that the error given by MIT is relatively large for the 9.60~G resonance.

Few theoretically weak resonances do not have an experimental counterpart, most likely since the corresponding experimental signature has been missed.
The quality of our assignment, yet non univocal, strongly suggests that the dominant anisotropic interaction arises from the electronic spins~\cite{1996-FHM-JRNIST-521}.

\subsection{Na$^{39}$K} \label{sec_nak39}
\begin{figure}[!hbt]
\centerline{\epsfig{file=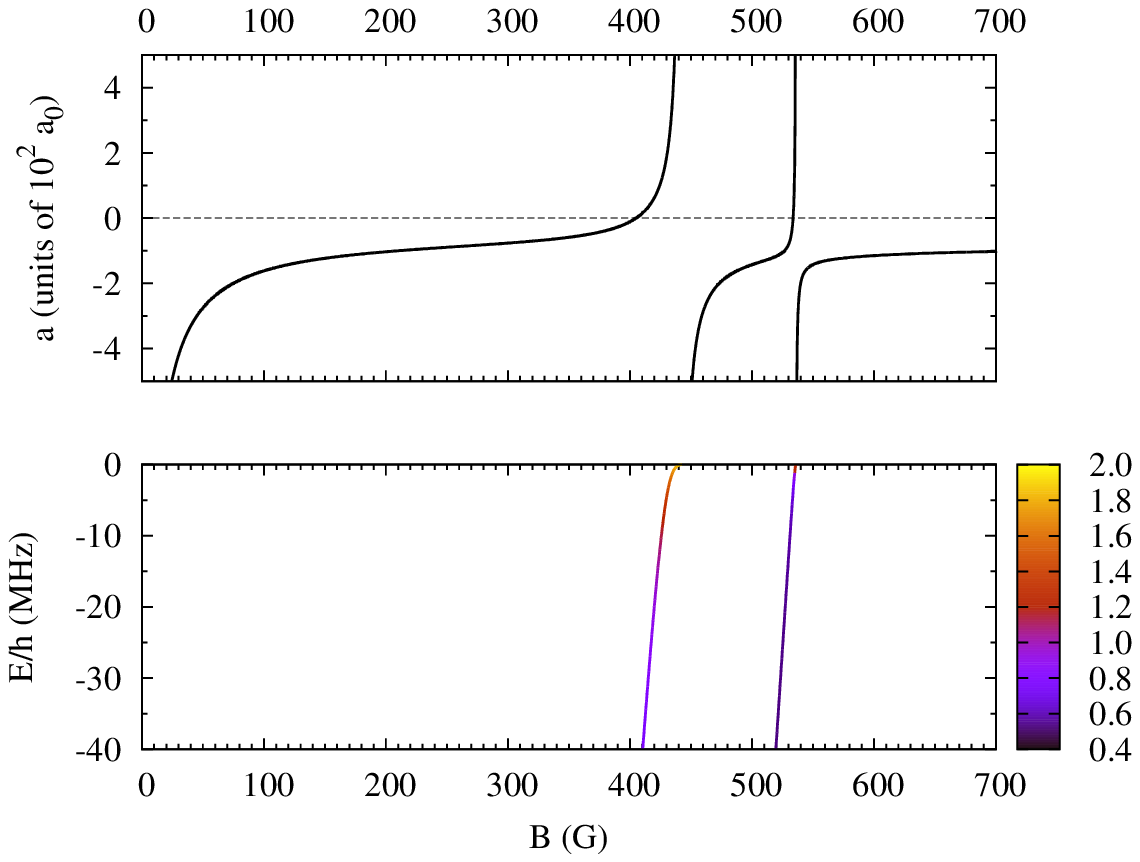,width=.99\columnwidth,angle=0}}
\caption[]{
(color online) Scattering length $a$ as a function of the magnetic field $B$
for Na$|1,1\rangle$ + $^{39}$K$|1 1 \rangle$ $s$-wave collisions (top panel).
Corresponding molecular energy levels as a function
of $B$ are shown in the lower panel. The density code denotes the average
spin $\langle {\vec S}^2 \rangle$ of the molecule. 
} \label{fig_nak39_aa} \end{figure}

\begin{figure}[!hbt]
\centerline{\epsfig{file=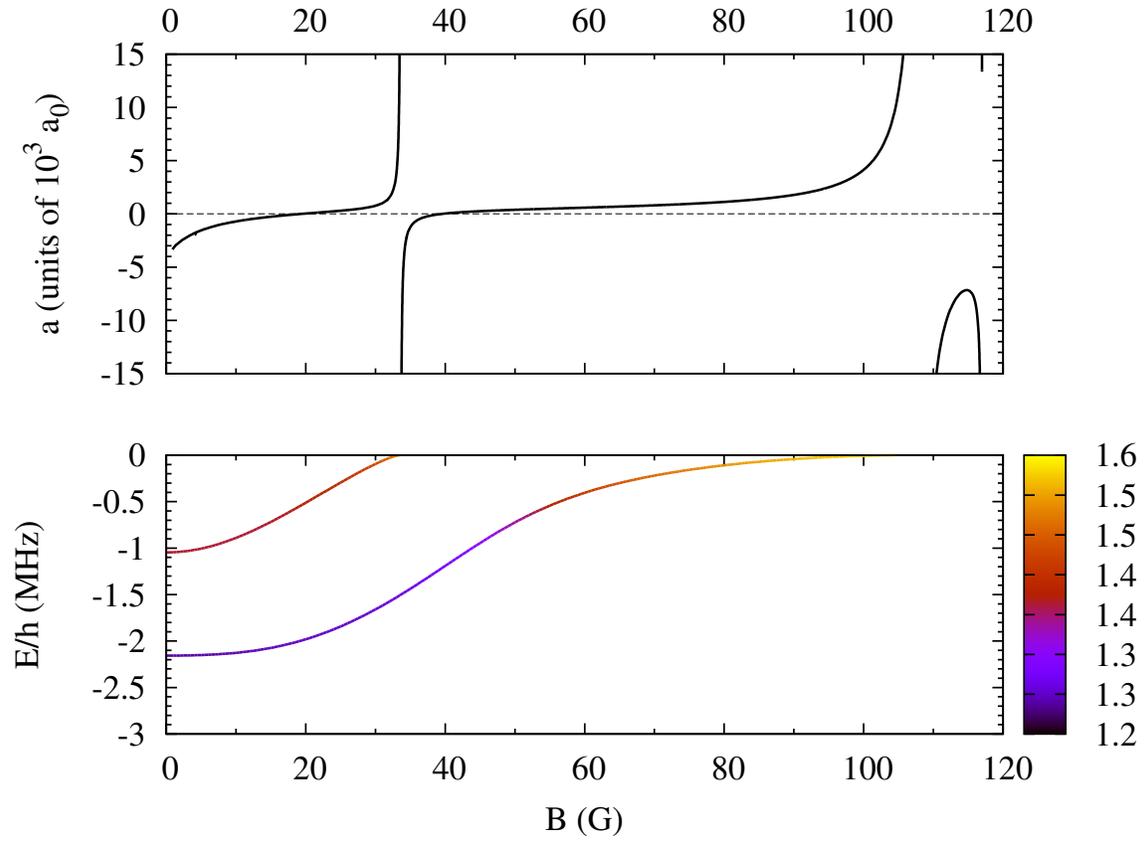,width=.99\columnwidth,angle=0}}
\caption[]{(color online) Same as Fig.~\ref{fig_nak39_aa} but for
for Na$|1,0\rangle$ + $^{39}$K$|1,0\rangle$ collisions.
} \label{fig_nak39_bb} \end{figure}

\begin{figure}[!hbt]
\centerline{\epsfig{file=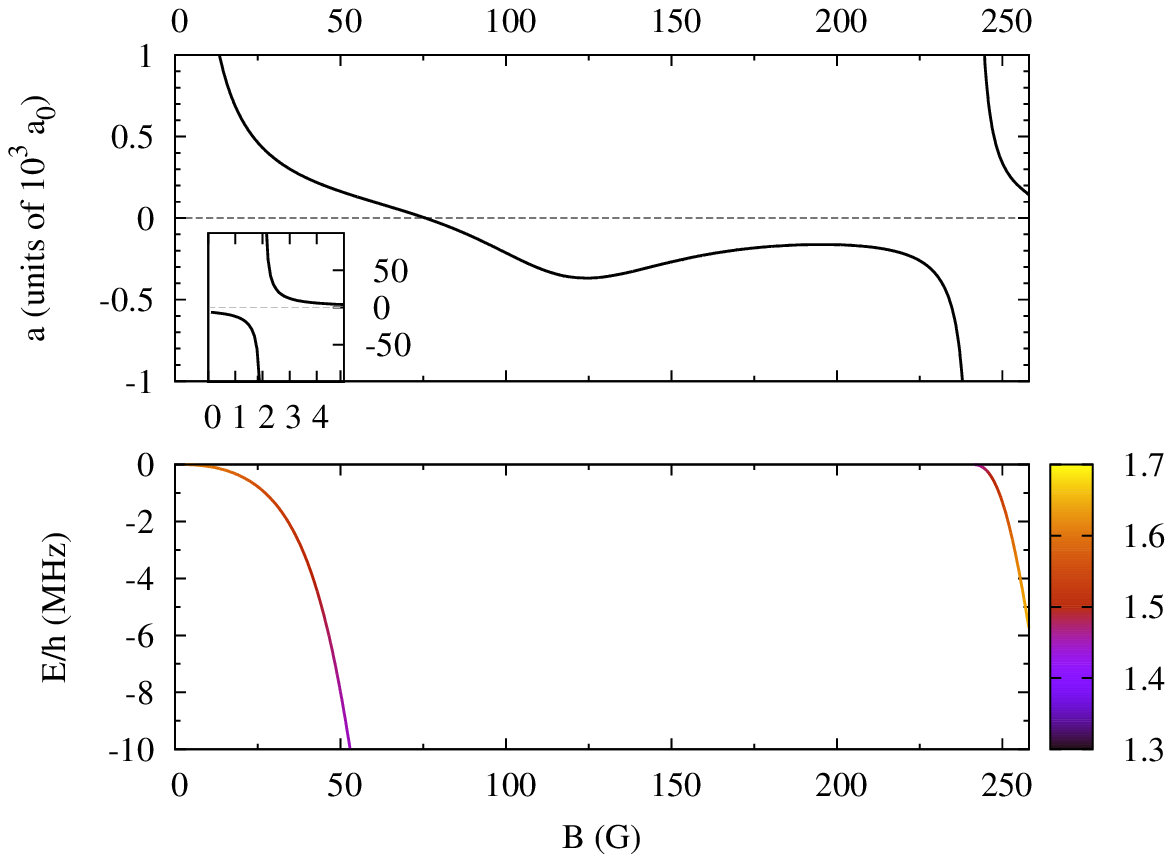,width=.99\columnwidth,angle=0}}
\caption[]{(color online) Same as Fig.~\ref{fig_nak39_aa} but for
for Na$|1,-1\rangle$ + $^{39}$K$|1,-1\rangle$ collisions. The inset shows the behavior of $a$ at low field
on a larger vertical scale.
} \label{fig_nak39_cc} \end{figure}

We continue our discussion with the most abundant potassium isotope,
$^{39}$K, a species for which cooling and Bose-Einstein condensation has
traditionally proved to be difficult, yet finally achieved by different
techniques~\cite{2012-ML-PRA-033421,PhysRevA.90.033405}. We provide
results for a series of hyperfine states. 
Our data can therefore be
useful in order to interpret collision data in a pure spin or in the
case of partial polarization of the sample. To this aim, calculations
of the $\mbox{$s$-wave}$ scattering length are performed for different values of
the conserved projection of total angular momentum, $M$. 
Tab.~\ref{tab_nak39} summarizes the $s$-wave
resonances found for Na$^{39}$K for a magnetic field up to 1000~G. We
report the positions of the 21 poles observed in the scattering length,
$B_\text{res}$, as well as the 17 zero-crossing field, $B_\text{ZC}$.
Note that no zero-crossing exists for Na $|1,0\rangle$ + $^{39}$K
$|1,-1\rangle$ collisions and that a single one at $B_\text{ZC}=75.71$~G
is present for the Na $|1,-1\rangle$ $^{39}$K $|1,-1\rangle$ channel.
\begin{table}[h!bt]
\caption[]{Same as Tab.~\ref{tab_nak40} but for the isotopic pair Na$^{39}$K. No experimental data are available for this system.
}
\label{tab_nak39}
\begin{ruledtabular}
\begin{tabular}{crrclrrr}
Na$^{39}$K channel                      & $B_\text{res}~(G)$ & $B_\text{ZC}~(G)$ & $r^\text{res}_\text{eff}~(a_0)$ &$s_\text{res}$ & $a_{bg}~(a_0)$ & $\varepsilon ~(G^{-1})$& $\Delta~(G)$  \\ \hline
$|1,1\rangle + |1,1\rangle$& 442.51 & 405.02 &   123.        &  4.00    &  -114.8 &3.72~10$^{-4}$ & -36.9      \\
                           & 536.00 & 533.72 &  -174.        &  0.316   &         &               & -2.27      \\ \hline
$|1,1\rangle + |1,0\rangle$&  35.16 &  11.51 &    62.3       &  1.18    &         &               &           \\
                           & 356.21 & 355.45 & -1.81~10$^3$  &  0.0520  &         &               &           \\
                           & 498.23 & 466.24 &   118.        &  3.35    &         &               &           \\
                           & 606.51 & 603.13 &   -88.2       &  0.430   &         &               &           \\ \hline
$|1,0\rangle + |1,0\rangle$&  33.60 &  19.39 & -1.60~10$^3$  &  0.0586  & 258.    &               &  8.2       \\
                           & 107.97 &  39.55 &   117.        &  3.22    & -569.    &              &  -62.      \\
                           & 116.91 &        & -969.         &  0.0912  & -10760  &               &  0.18      \\ \hline
$|1,1\rangle + |1,-1\rangle$& 116.98 &       & -3.85~10$^4$  & 0.00264  &         &               &            \\
                            & 422.51 & 421.86& -2.02~10$^3$  &  0.0469  &         &               &            \\
                           & 566.06 & 539.80 &   111.        &  2.74    &         &               &            \\
                           & 688.63 & 685.97 &  -170.        &  0.320   &         &               &           \\ \hline
$|1,-1\rangle +|1,0\rangle$&  56.31 &  54.70 &-3.04.~10$^4$  &  0.00334 &         &               &           \\ \hline
$|1,0\rangle + |1,-1\rangle$& 158.18 &       &   141.        & 14.1     &         &               &         \\ \hline
$|1,1\rangle + |2,-2\rangle$& 498.48 & 498.22& -4.93~10$^3$  &  0.0201  &         &               &         \\
                           & 648.26 & 627.58 &   102.        &  2.20    &         &               &         \\ \hline
$|1,-1\rangle +|1,-1\rangle$&   2.01 & 75.71 &   142.        & 15.2     & -183.7      &1.86~10$^{-2}$& 63.2       \\
                           & 241.40 &        &   107.        &  2.47    & -62.6       &           & 53.0         \\ \hline
$|1,0\rangle + |2,-2\rangle$& 357.96 &357.10 & -9.66~10$^3$  &  0.0104  &         &               &          \\
                           & 657.18 & 599.44 &   132.        &  6.30    &         &               &          \\
\end{tabular}
\end{ruledtabular}
\end{table}
Since we include only $s$-waves, possible narrow spin-spin resonances
due for instance to $s \to d$ wave couplings are not reproduced by the
model. The incoming state for the collision is systematically taken
to be the lowest energy state with the given $M$ at magnetic field
intensity $B$.  Note that in general this state may decay by inelastic
spin-spin processes if $\ell >0$ states were included in the basis. These
processes will however tend to be slow except very close to resonance and
are neglected for computational simplicity.  Figures \ref{fig_nak39_aa},
\ref{fig_nak39_bb} and  \ref{fig_nak39_cc} provide the scattering length
as well as the molecular energies for three of the nine studied channels.
Numerical data are available upon request to the authors.

As in the boson-fermion case we choose to parameterize the field-dependent
$s$-wave scattering length by the unique expression Eq.~\eqref{eq_a_width}
over a magnetic field range of $\pm 4\Delta$ around the resonance
and compute the resonance length in order to assess the resonance
strength. 
For overlapping resonances a unique $a_{bg}$ and
$\epsilon$ values are given, whereas for isolated resonances, we give a
local $a_{bg}$ and $\epsilon$, when the latter is non vanishing. We achieve
the sufficient required accuracy (below 5\% as for the $^{40}$K isotope)
for three of the nine initial channels considered.
Many combinations are found to be not well described
by Eq.~\eqref{eq_a_width}, in particular in the presence of
energetically degenerate channels that give rise to characteristic
threshold singularities~\cite{1959-JRN-PR-1611}.  

Both open $s_\text{res} \gg 1$ and closed $s_\text{res} \ll 1$
channel dominated resonances are available in suitable hyperfine
combinations. A particularly interesting feature is the one near 442~G
for collisions in the absolute hyperfine ground state $|f_a=1,m_a=1
\rangle + |f_a=1,m_a=1 \rangle$, which is strictly stable under
two-body inelastic collisions and open-channel dominated; See also
Figure~\ref{fig_nak39_aa}.  Its large magnetic width $|\Delta|=36.9$~G
should allow one to tune $a$ to desired values with high accuracy
and thus possibly to explore the quantum phases predicted in
free space and under optical-lattice confinement for a variety of
geometries~\cite{2003-LMD-PRL-090402,2003-EA-NJP-113,2014-MGM-NJP-103004}.
Also note that for vanishing $B$ the scattering length $a$ is negative
and very large in magnitude $|a|>5\times 10^3~a_0$, a feature related to
the presence of a virtual state with positive energy. Variation with $B$
of the position of the virtual state results in the rapid variation of
$a$ with magnetic field observed for small $B$.

The known FR for Na + Na collisions in the ground state are located
at large fields $B>800$~G~\cite{2002-TL-PRA-023412} in a region where
$a_{\rm KK}$ and $a_{\rm NaK}$ present regular non resonant
behavior~\footnote{For reference, $a_{\rm KK} \simeq -35 a_0$ and $a_{\rm NaK}
\simeq -90 a_0$.}. Tuning of the interspecies scattering length can be
used to increase the cross section for sympathetic cooling, for instance
to cool $^{39}$K by thermal contact with ultracold Na. A comparison
of Fig.~\ref{fig_nak39_aa} with the Fig.4 of Ref.~\cite{2007-CDE-NJP-223}
shows that at the field $B =395.2$~G at which $^{39}$K has been condensed
\cite{2007-GR-PRL-010403} the $a_{\text NaK}$ is slightly negative. A double
BEC of sodium and potassium will thus be miscible and stable against
collapse~\cite{PhysRevA.65.063614}. Moreover, if the double condensate
is adiabatically loaded in an optical lattice the attractive character
of the NaK effective interaction will favor the loading of Na and K
pairs at the lattice cells. This should be an advantageous starting
point to associate Feshbach molecules and thus implement STIRAP schemes
to form ultracold molecules in the absolute ground state. 

\begin{figure}[!hbt]
\centerline{\epsfig{file=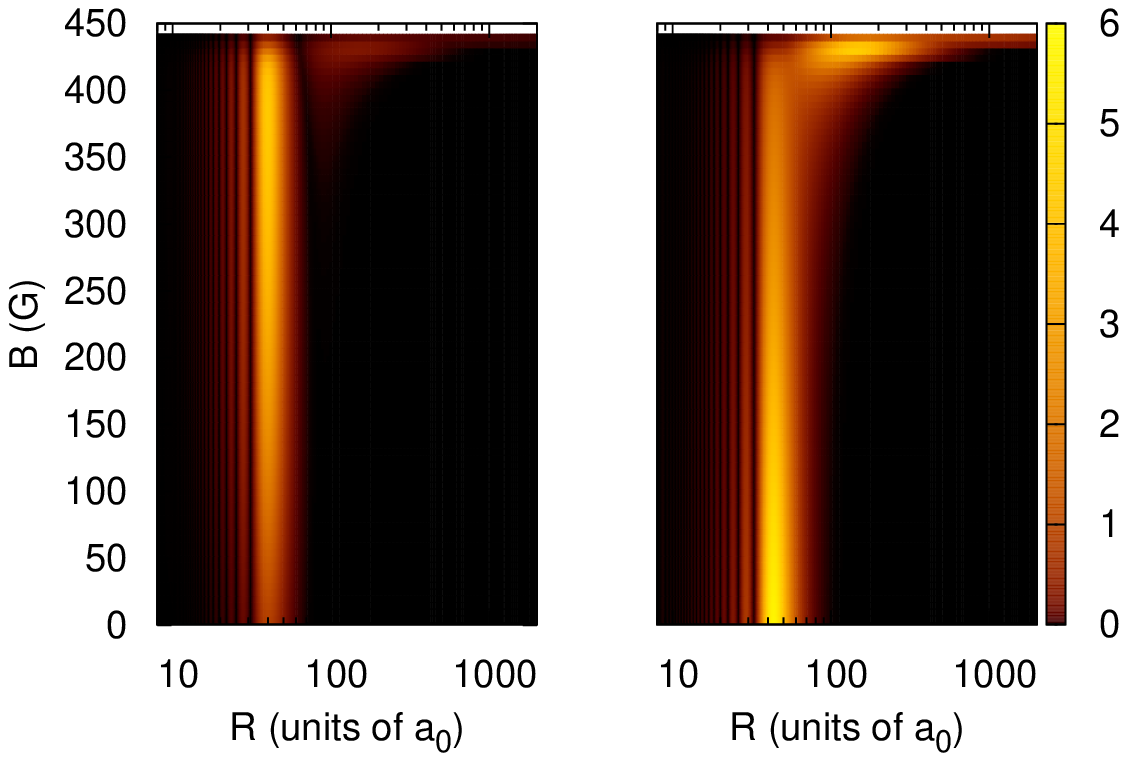,width=.99\columnwidth,angle=0}}
\caption[]{(color online) The $S=0$ (left
panel) and $S=1$ (right panel) electron spin components in arbitraty units of last-below-threshold $\ell=0$ multichannel wavefunction 
with $M=2$ as a function of the magnetic field $B$ for the Na$^{39}$K dimer. 
}
\label{fig_nak39_aa_wave_1}
\end{figure}

Indeed, as compared to the boson-fermion case of Sec.~\ref{sec_nak40} one
can verify from the $\langle {\vec S}^2 \rangle$ given by the color code
in the lower panel of Fig.~\ref{fig_nak39_aa} that the situation is here
favorable, since the molecule presents hyperfine-induced singlet-triplet
mixing even far from dissociation. Beyond the average spin character,
we also represent in Fig.~\ref{fig_nak39_aa_wave_1} the detail of the
singlet and triplet components of the coupled wave-function, defined
as $\Psi_{0} = {\hat P}_{0} \Psi$ and $\Psi_{1} = {\hat P}_{1} \Psi$
with ${\hat P}_{0,1}$ the projectors on the $S=0$ and $1$ subspaces,
respectively.
Interestingly, the $S=0$ amplitude reaches its maximum right
before the resonance, at $B \approx 400$~G. Most importantly,
Fig.~\ref{fig_nak39_aa_wave_1} shows that $\Psi_0$ maintains a short-range
character with maximum amplitude for $\mbox{$R \approx 40 ~ a_0$.}$

\begin{figure}[!hbt]
\centerline{\epsfig{file=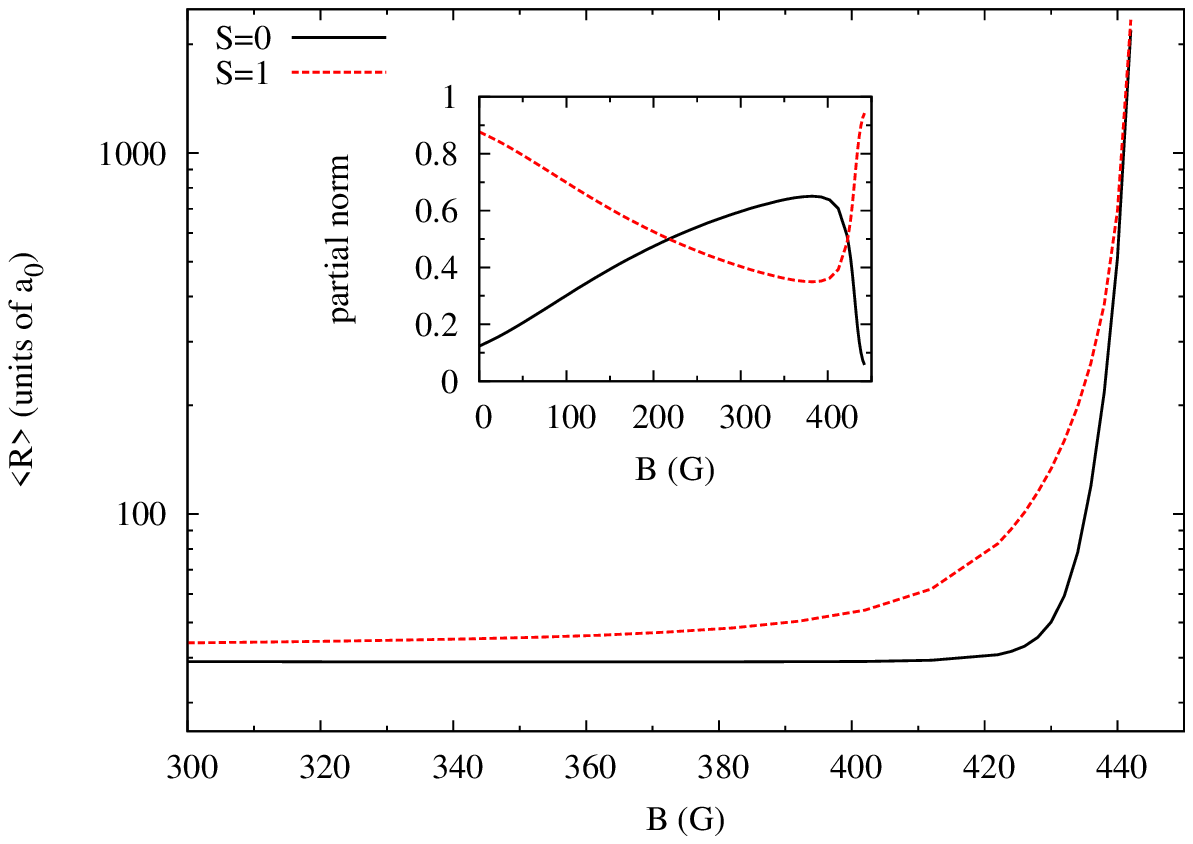,width=.99\columnwidth,angle=0}}
\caption[]{Average distance $\langle R \rangle$ and partial singlet/triplet norms (defined in the text) of the last-below-threshold $\ell=0$ multichannel wavefunction with $M=2$ as a function of the magnetic field $B$ for the Na$^{39}$K dimer.  }
\label{fig_nak39_aa_averager_1}
\end{figure}

Such short-range character is confirmed quantitatively in
Fig.~\ref{fig_nak39_aa_averager_1} by calculations of the average distances
$\langle R \rangle_{0,1} = \langle \Psi_{0,1} | {\hat R} | \Psi_{0,1}
\rangle $. 
Moreover, the partial norms $|\langle \Psi_{0,1} | \Psi_{0,1}
\rangle |^2 $ depicted in the figure show that the singlet admixture is
significant in the region of interest. Analyses of the electronic excited state 
structure of NaK and of the corresponding Frank-Condon factor for transfer of
the Feshbach molecule to the excited state is beyond the scope of this
work. However Ref.~\cite{2013-TAS-PRA-023401} finds relatively favorable Frank-Condon
factors in the case of Bose-Fermi Feshbach molecules, that present similar
spatial extent as the present bosonic ones but with significantly smaller
singlet component. We expect therefore that suitable excited states can be
found to implement an efficient two-photon transfer in the present case.

Let us now consider collisions for atoms in the first excited hyperfine
level $|1,0\rangle$ reproduced in Fig.~\ref{fig_nak39_bb}.  Similar to
the case of the absolute ground state, a large negative scattering length
rapidly varying with $B$ is predicted at low magnetic fields.  A point
of non-analiticity is expected at a magnetic field $B\simeq 117$~G
as the $|1,0\rangle + |1,0\rangle$ and $|1,1\rangle + |1,-1 \rangle$
channels become degenerate. It is interesting to observe here that the
expected cusp in the elastic scattering matrix element is accompanied
by poles in $a$ occurring right before (after) the degeneracy point
in the $|1,0\rangle + |1,0\rangle$ (the $|1,1\rangle + |1,-1 \rangle$)
channel a peculiar effect stemming from the interplay beetween channel
degeneracy and Feshbach physics; see Fig.~\ref{fig_nak39_bb} and
Tab.~\ref{tab_nak39}.

The case of $|1,-1\rangle + |1,-1\rangle$ collisions is shown in
Fig.~\ref{fig_nak39_cc} and is particularly relevant for the applications
since $|1,-1\rangle$ is the lowest magnetically trappable atomic state of
$^{39}$K and Na at low $B$.  Moreover, below 259~G the NaK $|1,-1\rangle
+ |1,-1\rangle$ combination is stable under $s$-wave collisions since it is the
lowest hyperfine state with $m_f=-2$.  Note that due to the presence of
a Feshbach resonance at very low magnetic fields $B=2~$G the scattering
zero-field length is negative and very large in magnitude (see inset
of Fig.~\ref{fig_nak39_cc}).
Bose-Einstein condensation has been achieved in this
hyperfine level using magnetic tuning of $a$ to suitable
values~\cite{2012-ML-PRA-033421,PhysRevA.90.033405}.  Interestingly,
Fig.~5 of Ref.~\cite{2007-CDE-NJP-223} shows that for magnetic field
between the two homonuclear K resonances located in $|1,-1\rangle$
at about 33 and 163~G the $a_{\rm KK}$ is positive, thus ensuring the
stability of a K condensate. In the same magnetic field region $a_{\rm
NaK}$ varies from being large and positive to large and negative,
allowing one to explore the phase diagram of a quantum degenerate NaK
mixture as a function of the mutual interaction strength.

\begin{figure}[!hbt]
\centerline{\epsfig{file=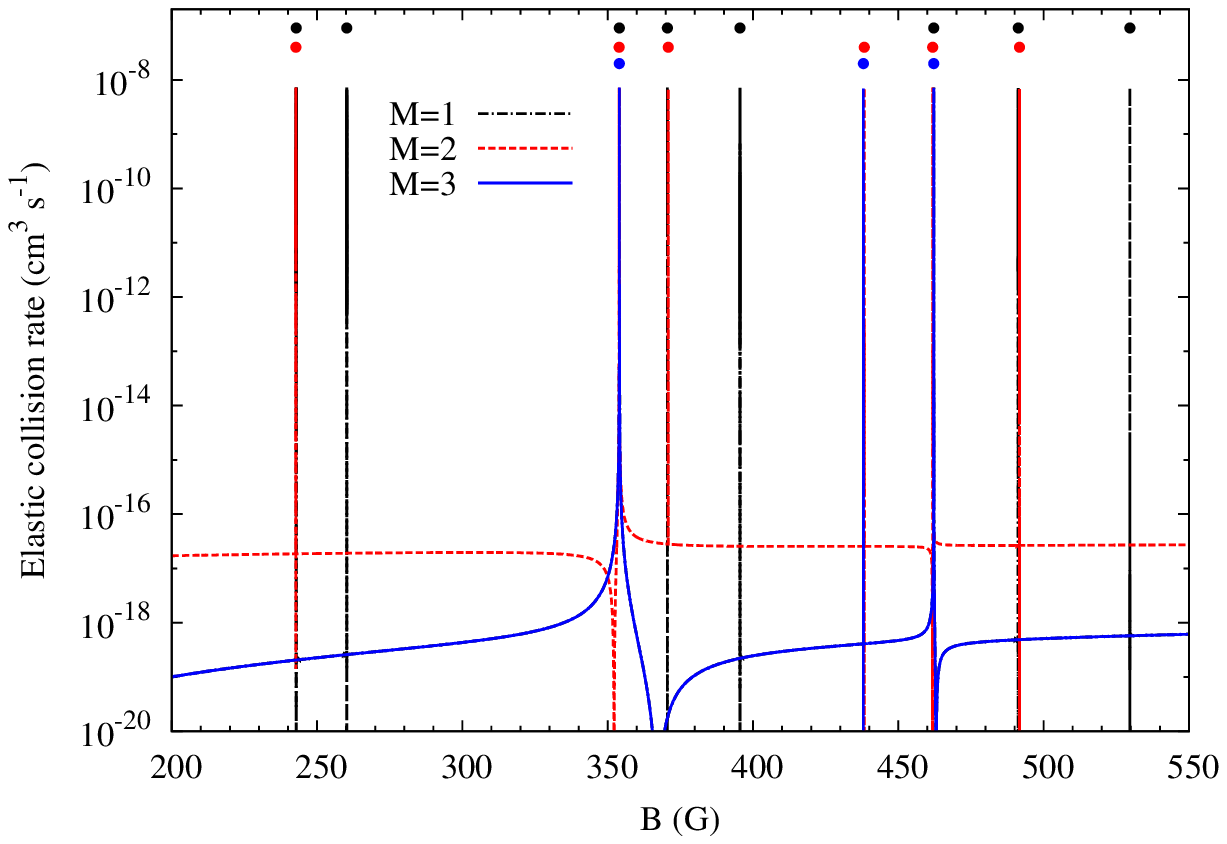,width=.99\columnwidth,angle=0}}
\caption[]{(color online) Elastic collision rate computed at a collision energy of $1~\mu$K as a function of the magnetic field $B$
for Na$|1,1\rangle$ + $^{39}$K$|1,1\rangle$ $p$-wave collisions.
The dots materialize the occurence of maxima in the rate for each projection $M$. 
} \label{fig_nak39_aa_p}
\end{figure}

To conclude our analysis for this isotope, we provide in Tab.~\ref{tab_nak39_p} the
spectrum of $p$-wave resonances, limiting ourselves to the absolute
ground state. 
\begin{table}[hb!t]
\caption[]{Theoretical magnetic field locations $B_\text{res}$ of $p$-wave
FR for Na$^{39}$K for projection of total angular momentum
$M$ in different hyperfine atomic channels. Calculations are performed at collision energy of $1\mu$K.}
\label{tab_nak39_p}
\begin{tabular}{p{7cm}ccp{7cm}}
\begin{ruledtabular}
\begin{tabular}{ccr}
Na$^{39}$K channel &$M$ & $B_\text{res}~(G)$  \\  \hline
$|1,1\rangle + |1,1\rangle$ & 1 & 242.83 	 \\
                            & 1 & 260.19 	 \\
                            & 1 & 354.02 	 \\
                            & 1 & 370.58 	 \\
                            & 1 & 395.53 	 \\
                            & 1 & 462.21 	 \\
                            & 1 & 491.27 	 \\
                            & 1 & 529.67 	 \\
&   &          \\
\end{tabular}
\end{ruledtabular}
&~~~~~ & &
\begin{ruledtabular}
\begin{tabular}{ccr}
Na$^{39}$K channel & $M$ & $B_\text{res}~(G)$  \\ \hline
$|1,1\rangle + |1,1\rangle$ &2 & 242.68       \\
                            &2 & 353.95       \\
                            &2 & 370.82       \\
                            &2 & 438.27       \\
                            &2 & 461.84       \\
                            &2 & 491.68       \\ \hline
$|1,1\rangle + |1,1\rangle$ &3 & 354.01       \\
                            &3 & 437.99       \\
                            &3 & 462.21       \\
\end{tabular}
\end{ruledtabular}
\end{tabular}
\end{table}
As in the case of the boson-fermion mixtures~\cite{2012-CHW-PRL-085301}
such resonances can be experimentally observable even at ultracold
temperatures. Fig.~\ref{fig_nak39_aa_p} shows the elastic collision rate for different $M$ 
projections presenting a rich spectrum with
nearby peaks of multiplicities three, two and one. Closer inspection shows that
triply degenerate peaks are the usual doublets~\cite{PhysRevA.69.042712}, with the
peaks arising from spin-spin induced mixing of $m_f=2, m=\pm 1$ states being
nearly degenerate and slightly shifted with respect to the $m_f=2, m=0$
peak. Larger multiplicities like the ones in Fig.~\ref{fig_nak40_aa_p} are not observed
here since $p$-wave resonances occur at larger magnetic fields; See
the discussion in Sec.~\ref{sec_nak40}. According to the value
of $M$ in Fig.~\ref{fig_nak39_aa_p}, doubly degenerate components arise
from coupling to states with $m_f=1, m=0,1$ or to states with $m_f=3,
m=-1,0$. Finally, all singly degenerate levels in the figure are due to
coupling with $m_f=0, m=1$.

\subsection{Na$^{41}$K} \label{sec_nak41}
\begin{figure}[!hbt]
\centerline{\epsfig{file=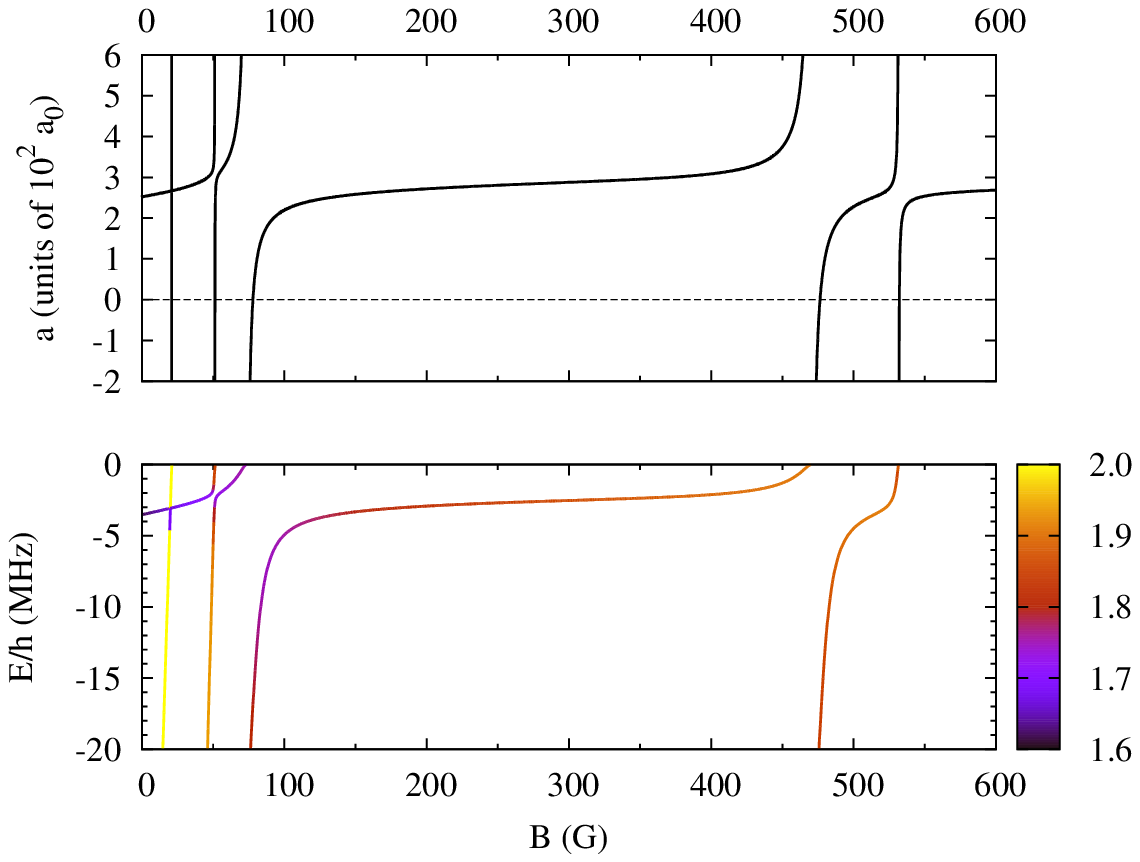,width=.99\columnwidth,angle=0}}
\caption[]{(color online) Scattering length $a$ as a function of the magnetic field $B$ for Na$|1,1\rangle$ + $^{41}$K$|1,1\rangle$ collisions (top panel) in the $s$-wave. Corresponding molecular energy levels as a function of $B$ are
shown in the lower panel.
The density code denotes the average spin $\langle {\vec S}^2 \rangle$ of the molecule. }
\label{fig_nak41_aa}
\end{figure}

We now provide numerical data for the other bosonic pair Na$^{41}$K.
Let us first recall that $^{41}$K has been brought to Bose-Einstein
condensation using Rb as a coolant or more recently by direct
evaporation~\cite{PhysRevA.79.031602}. Resonances exist for collisions
in different hyperfine states with magnetic widths of several Gauss; see
Tab.~\ref{tab_nak41}. 
\begin{table}[hb!t]
\caption[]{Same as Tab.~\ref{tab_nak39} but for Na$^{41}$K.}
\label{tab_nak41}
\begin{ruledtabular}
\begin{tabular}{crrclrrc}
Na$^{41}$K channel                      & $B_\text{res}~(G)$ & $B_\text{ZC}~(G)$ & $r^\text{res}_\text{eff}~(a_0)$ & $s_\text{res}$& $a_{bg}~(a_0)$ & $\varepsilon ~(G^{-1})$& $\Delta~(G)$  \\ \hline
$|1,1\rangle + |1,1\rangle$  &  20.90 &  20.90 &   -6.55~10$^{6}$&  1.56~10$^{-5}$ & 334.80 & -1.03~10$^{-4}$ & 3.57~10$^{-5}$ \\
                             &  51.23 &  51.30 &   -2.76~10$^{3}$&  3.52~10$^{-2}$ &        &                 & 7.10~10$^{-2}$ \\
                             &  73.35 &  77.97 &   85.2          &  1.60           &        &                 & 4.59          \\
                             & 470.08 & 476.41 &  104.           &  2.25           &        &                 & 6.32              \\
                             & 531.59 & 532.16 & -458.           &  1.68~10$^{-1}$ &        &                 & 5.63~10$^{-1}$ \\
                             &(-235.65)&       &                 &                 &        &                 &(6.89~10$^{1}$)\\ \hline
$|1,1\rangle + |1,0\rangle$  &  33.26 &  33.26 &  -6.18~10$^{5}$ &  1.66~10$^{-4}$ &        &                 &                \\
                             &  35.53 &  35.53 &  -3.70~10$^{7}$ &  2.76~10$^{-6}$ &        &                 &                 \\
                             &  66.48 &  66.61 &  -1.58~10$^{3}$ &  5.93~10$^{-2}$ &        &                 &                 \\
                             &  87.53 &  90.94 &  58.1           &  1.12           &        &                 &                \\ 
                             & 165.58 & 165.60 &  -4.72~10$^{4}$ &  2.16~10$^{-3}$ &        &                 &                 \\
                             & 453.37 & 453.37 &  -1.03~10$^{5}$ &  9.88~10$^{-4}$ &        &                 &                 \\
                             & 499.41 & 506.39 & 108.            &  2.50           &        &                 &                  \\
                             & 566.30 & 567.17 &-244.            &  2.60~10$^{-1}$ &        &                 &                 \\ \hline
$|1,0\rangle + |1,0\rangle$  &  35.05 &  35.05 &  -5.67~10$^{6}$ &  1.80~10$^{-5}$ & 246.1  & 9.71~10$^{-4}$  & 8.08~10$^{-4}$  \\ \hline
$|1,1\rangle + |1,-1\rangle$ &  63.46 &  63.48 &  -2.87~10$^{4}$ &  3.55~10$^{-3}$ &        &                 &                 \\
                             &  72.53 &  72.53 &  -8.85~10$^{5}$ &  1.16~10$^{-4}$ &        &                 &                  \\
                             & 106.20 & 107.69 &  -77.5          &  4.51~10$^{-1}$ &        &                 &                \\
                             & 183.36 & 183.36 &   -1.25~10$^{7}$&  8.17~10$^{-6}$ &        &                 &                  \\
                             & 370.11 & 370.11 &   -1.20~10$^{6}$&  8.51~10$^{-5}$ &        &                 &                  \\
                             & 481.53 & 481.53 &   -1.34~10$^{4}$&  7.62~10$^{-4}$ &        &                 &                  \\
                             & 531.87 & 539.42 &  111.           &  2.70           &        &                 &                  \\
                             & 604.62 & 605.36 & -309.           &  2.23~10$^{-1}$ &        &                 &                  \\ \hline
$|1,0\rangle + |1,-1\rangle$ &  66.97 &  66.98 &   -8.54~10$^{5}$&  1.20~10$^{-4}$ &        &                 &                   \\
                             & 129.37 & 129.60 &   -1.98~10$^{3}$&  4.81~10$^{-2}$ &        &                 &                 \\
                             & 149.33 & 149.32 &   -2.68~10$^{5}$&  3.82~10$^{-4}$ &       &                  &                 \\ \hline
$|1,1\rangle + |2,-2\rangle$ & 156.22 & 156.22 &   -1.16~10$^{6}$&  8.84~10$^{-5}$ &        &                 &                  \\
                             & 209.92 & 209.93 &   -5.51~10$^{4}$&  1.85~10$^{-3}$ &        &                 &                  \\
                             & 391.25 & 391.25 &   -1.26~10$^{6}$&  8.13~10$^{-5}$ &        &                 &                  \\
                             & 512.63 & 512.63 &   -3.84~10$^{5}$&  2.66~10$^{-4}$ &        &                 &                  \\
                             & 567.79 & 575.74 &  113.           &  2.82           &        &                 &                    \\ \hline
$|1,-1\rangle + |1,-1\rangle$ & 137.27 & 137.27&   -4.13~10$^{5}$&  2.48~10$^{-4}$ & 212.75& 0               & -2.43~10$^{-3}$   \\ \hline
$|1,0\rangle + |2,-2\rangle$ & 146.65 & 146.65 &   -9.73~10$^{5}$&  1.05~10$^{-4}$ &        &                 &                  \\
                             & 245.19 & 252.66 &   91.4          &  1.77           &        &                 &       \\
                             & 500.76 & 500.82 &   -9.51~10$^{3}$&  1.06~10$^{-2}$ &        &                 &                  \\
                             & 601.15 & 606.56 &   92.9          &  1.81           &        &                 &        \\ \hline
\end{tabular}
\end{ruledtabular}
\end{table}
Such broad resonances are essentially open-channel
dominated, with resonance strength of $s_{\rm res} \lesssim 3$. Several closed channel
dominated features are also readily available in each hyperfine channel we studied. A
distinctive feature of Na$^{41}$K is the large and positive $a_{bg}$
for all the hyperfine combinations.
The parameterization Eq.~\eqref{eq_a_width} for overlapping resonances is
used for the absolute ground state, where it is found to be accurate only
if an artificial pole is added in  Eq.~(\ref{eq_a_width}) at negative
$B$. Such a pole mimicks the effect of a virtual state, {\it i.e.}
a quasi-bound state located at positive energy and that would give a
resonance at negative values of $B$. The position obtained through the
fitting procedure is given in Table~\ref{tab_nak41} in parenthesis to
distinguish from physical poles of $a$.
The corresponding scattering length $a$ is given in the top panel of Fig.~\ref{fig_nak41_aa}.

Both $^{41}$K and Na homonuclear resonances in $|1, 1 \rangle + |1, 1
\rangle$ are narrow and quite sparse. Combination of the present and the
magnetic spectra in Refs~\cite{2007-CDE-NJP-223} and~\cite{2002-TL-PRA-023412}
for K and Na respectively shows that homonuclear and heteronuclear
resonances take place at well separated locations.  Note that the large
$a_{bg}$ for NaK and the nonresonant values of order $50a_0$ for both Na
and $^{41}$K imply that two Bose-Einstein condensates will tend to phase
separate. However, the heteronuclear resonances can be used to reduce
or even change the sign of $a_{\rm NaK}$, such to favor miscibility and
eventually the realization of overlapping quantum gases of Na and K in free
space or in optical lattices.

Let us now discuss magnetic association of Na and K atoms when they
are prepared in the respective ground hyperfine levels. A calculation
of the quantum average $\langle \vec S \rangle^2$ depicted as density
code in the lower panel of Fig.~\ref{fig_nak41_aa} readily shows that
resonances arise from states with dominant triplet character. Note
that the large background scattering length implies the existence
of a molecular level close to the dissociation threshold; See lower
panel of Fig.~\ref{fig_nak41_aa}. Let us consider performing magnetic
association near the two broadest FR. Based on our data, three routes
can be envisioned, yet presenting drawbacks.

\begin{figure}[!hbt]
\centerline{\epsfig{file=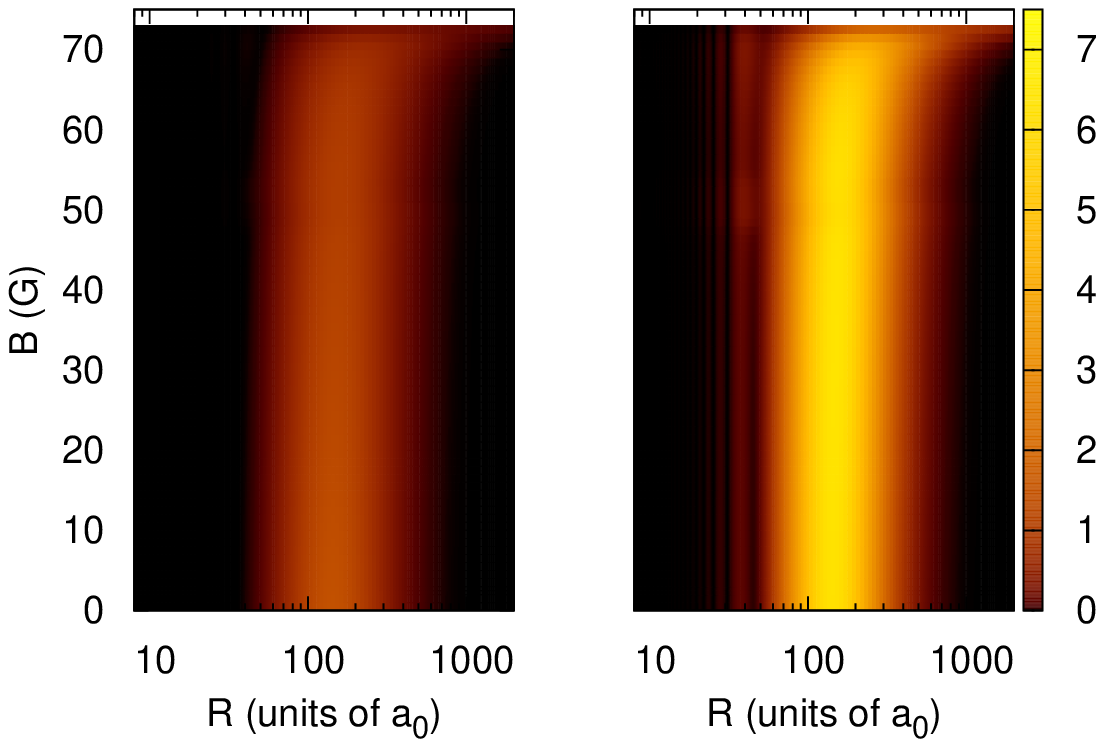,width=.99\columnwidth,angle=0}}
\caption[]{ (color online) The $S=0$ (left panel) and $S=1$ (right panel) electron spin components using the same arbitrary units of fig~\ref{fig_nak39_aa_wave_1} of the $\ell=0$ multichannel wavefunction with $M=2$ as a function of the magnetic field $B$ for the Na$^{41}$K dimer. The molecule is created at 73.35~G and the molecular state is followed diabatically with decreasing $B$.}
\label{fig_nak41_aa_wave_1}
\end{figure}

\begin{figure}[!hbt]
\centerline{\epsfig{file=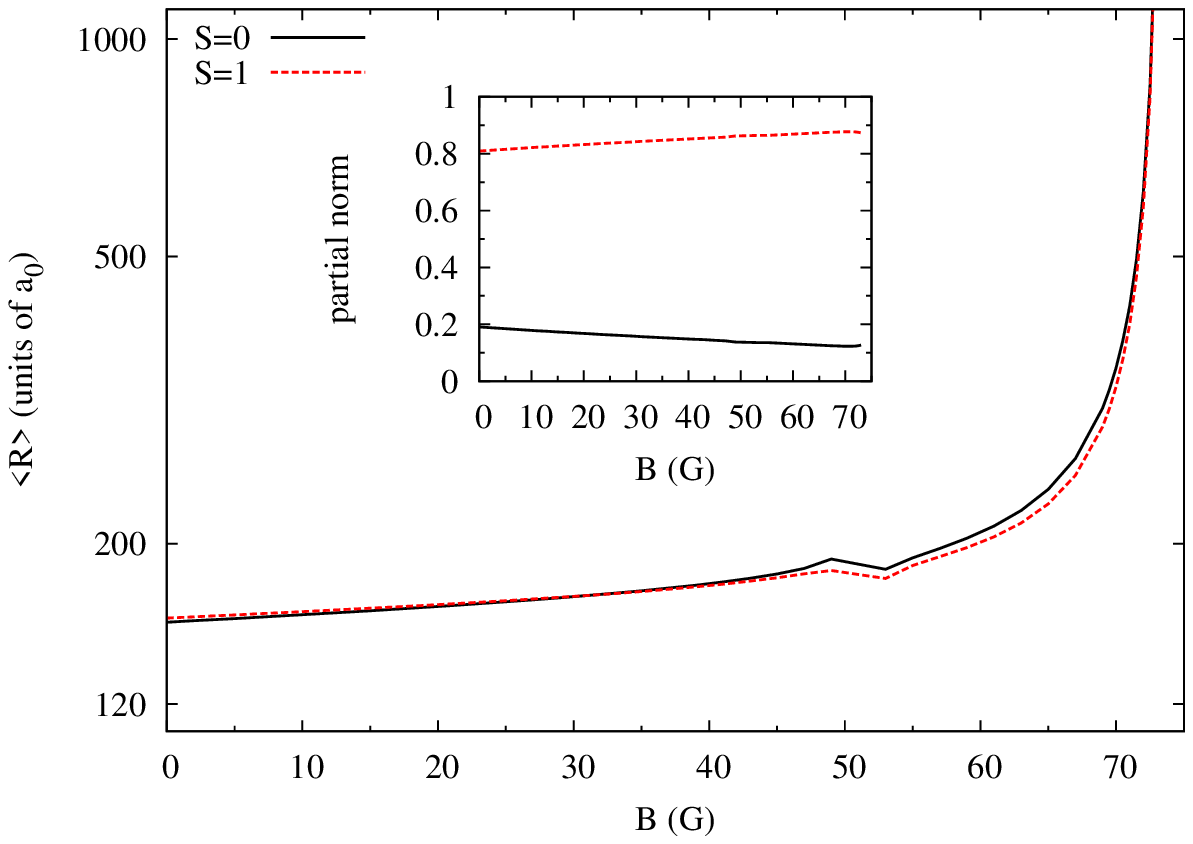,width=.99\columnwidth,angle=0}}
\caption[]{ Average distance $\langle R \rangle$ and partial singlet/triplet norms (defined in the text) of the last-below-threshold $\ell=0$ multichannel wavefunction with $M=2$ as a function of the magnetic field $B$ for the Na$^{39}$K dimer.  The molecular state is followed diabatically as for Fig.~\ref{fig_nak41_aa_wave_1}.  }
\label{fig_nak41_aa_averager_1}
\end{figure}

If molecules are formed at the 73~G FR and molecular curve crossings
are swept through diabatically, one stays in the ``background''
weakliest bound molecular level. Unfortunately, as shown in
Fig.~\ref{fig_nak41_aa_wave_1} and~\ref{fig_nak41_aa_averager_1}, the
state has long-range character with $\langle R \rangle_{0,1} \gtrsim 100~a_0$. 
Therefore, in spite of the sufficient singlet character predicted in
Fig.~\ref{fig_nak41_aa_averager_1} poor overlap is expected with the
excited molecular states. Note that since quantum numbers of this
background state are essentially atomic ones or Hund's case (e) the
projections $\Psi_0$ and $\Psi_1$ on the Hund's case (b) spin-coupled
basis in Fig.~\ref{fig_nak41_aa_wave_1} have virtually identical spatial
profiles.

An alternative route consists in following adiabatically the entrance
state through the first avoided crossing near 50~G. As shown in the lower
panel of Fig.~\ref{fig_nak41_aa} this leads however to the formation of
a molecule with relatively poor singlet admixture.

\begin{figure}[!hbt]
\centerline{\epsfig{file=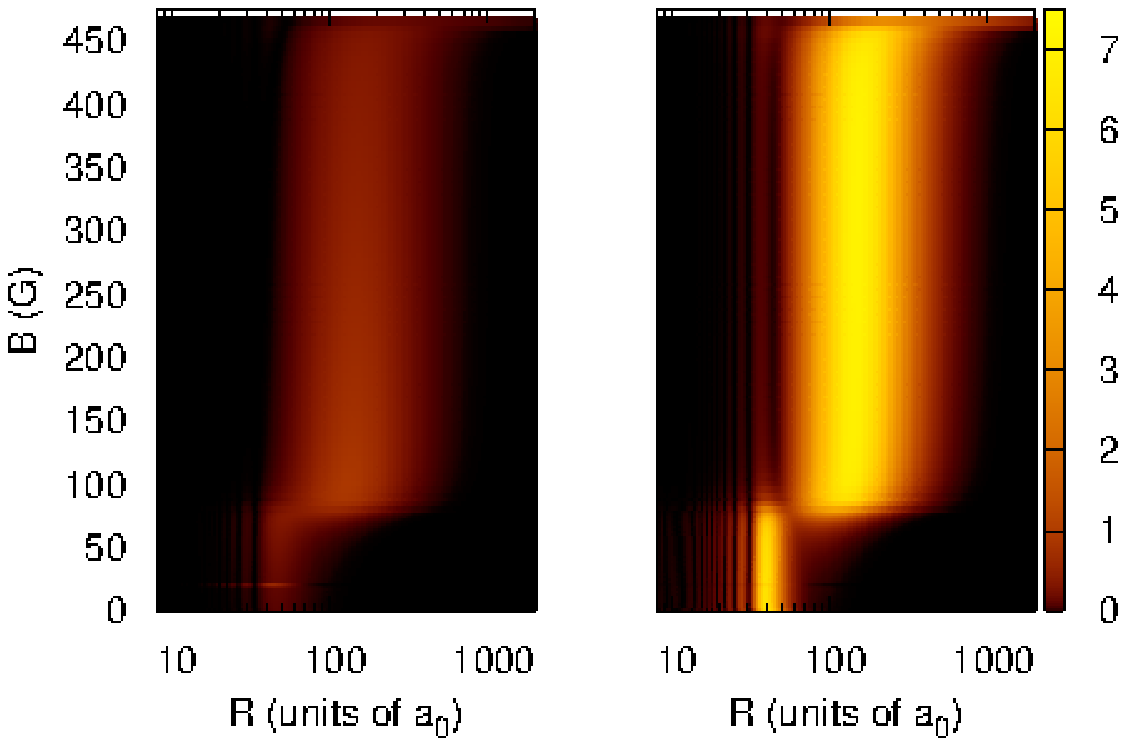,width=.99\columnwidth,angle=0}}
\caption[]{ (color online) The $S=0$ (left panel) and $S=1$ (right panel) electron spin components using the same arbitrary units of fig~\ref{fig_nak39_aa_wave_1} of the $\ell=0$ multichannel wavefunction with $M=2$ as a function of the magnetic field $B$ for the Na$^{41}$K dimer. The molecule is created at 470~G and the molecular state is followed adiabatically with decreasing $B$.  }
\label{fig_nak41_aa_wave_2}
\end{figure}

\begin{figure}[!hbt]
\centerline{\epsfig{file=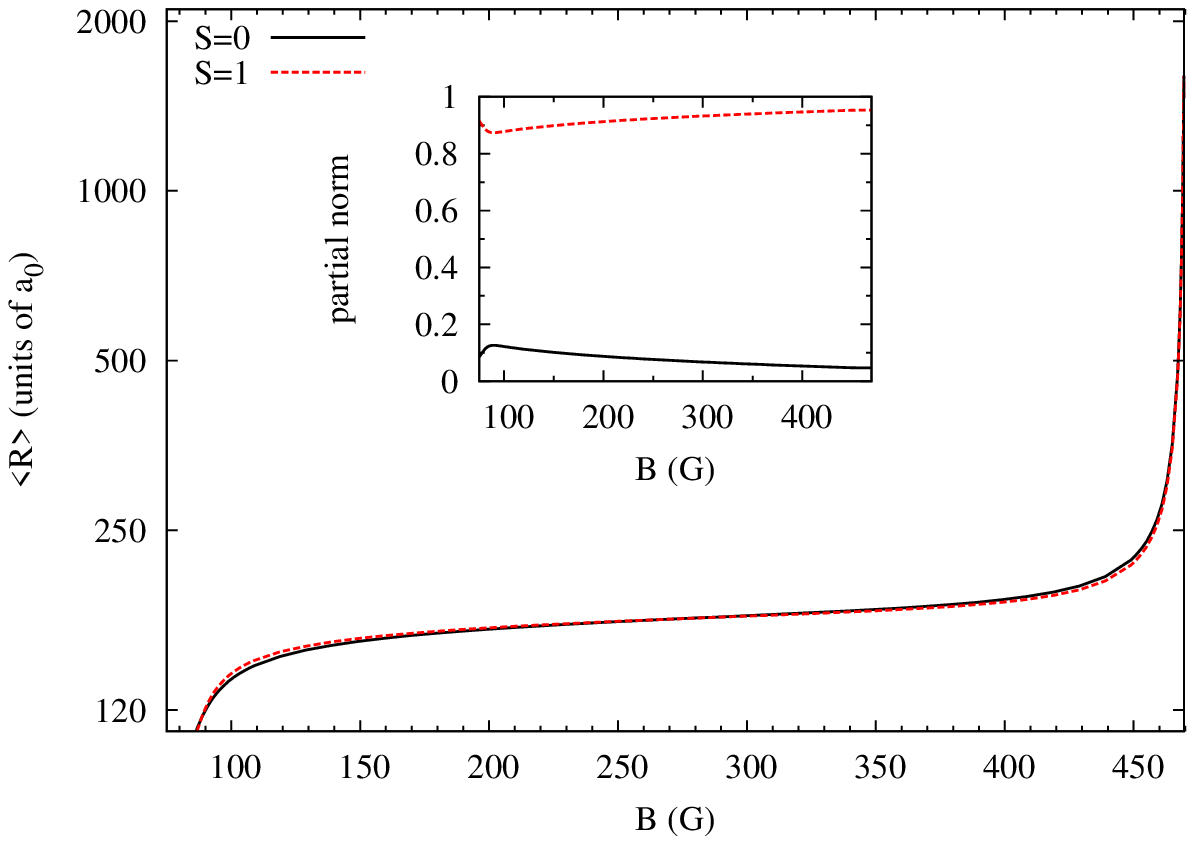,width=.99\columnwidth,angle=0}}
\caption[]{ Average distance $\langle R \rangle$ and partial singlet/triplet norms (defined in the text) of the last-below-threshold $\ell=0$ multichannel wavefunction with $M=2$ as a function of the magnetic field $B$ for the Na$^{39}$K dimer.  The molecular state is followed adiabatically as for Fig.~\ref{fig_nak41_aa_wave_2}.  }
\label{fig_nak41_aa_averager_2}
\end{figure}

If one uses the broad resonance at 470~G as an entrance gate,
a long magnetic field sweep down to $B\sim60$~G would be needed
before a small $\langle R \rangle_0$ is attained, as it can be
inferred from Fig.~\ref{fig_nak41_aa_wave_2} and the main panel of
Fig.~\ref{fig_nak41_aa_averager_2}. 
However, the inset in the latter
figure indicates that molecule shrinking also corresponds to a
drop in the singlet character, thus requiring a compromise to be found.
In order to draw firmer conclusions a detailed analysis of the excited
states will be needed.

\begin{figure}[!hbt]
\centerline{\epsfig{file=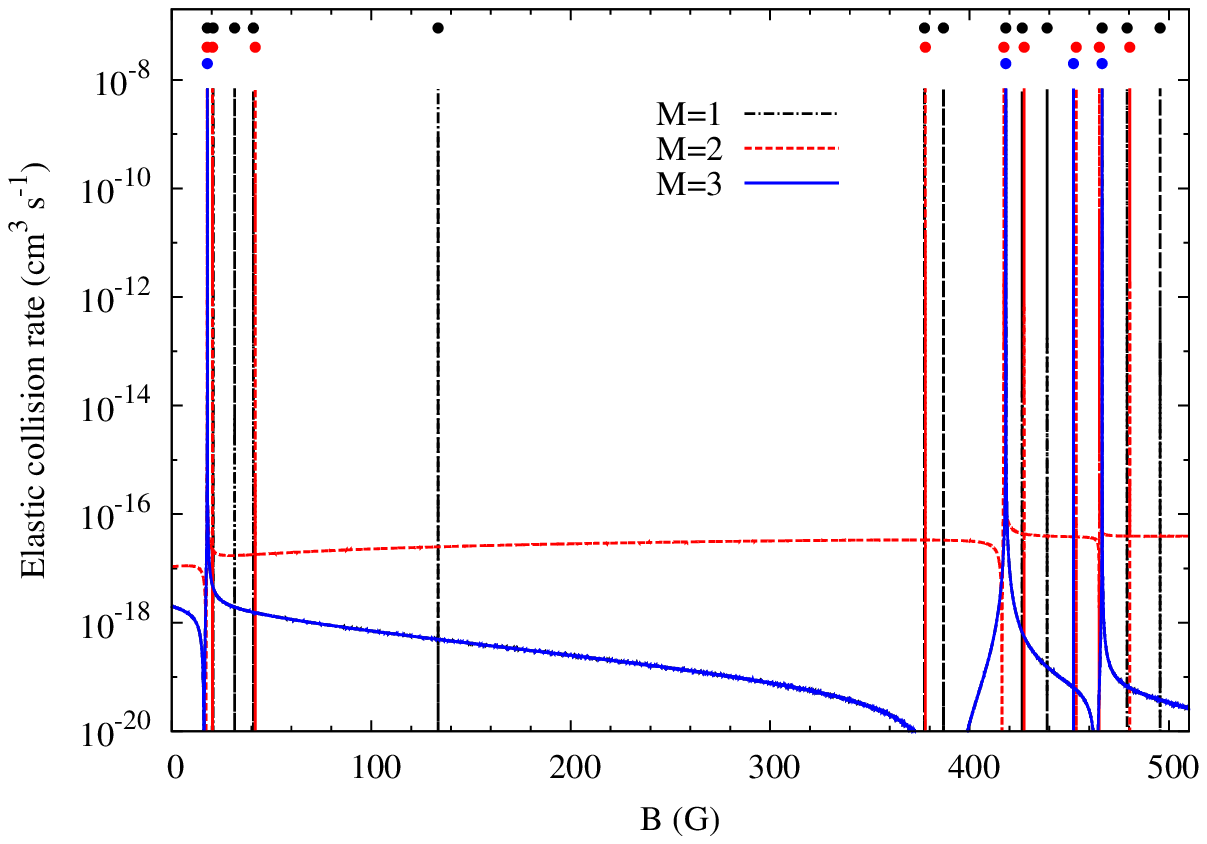,width=.99\columnwidth,angle=0}}
\caption[]{(color online) Elastic rate computed at a collision energy of $1~\mu$K as a function of the magnetic field $B$ for Na$|1,1\rangle$ + $^{41}$K$|1,1\rangle$ $p$-wave collisions.  The color dots above each peaks materialize the occurence of maxima in the rate for each projection $M$.  } 
\label{fig_nak41_aa_p}
\end{figure}

Finally, as illustrated in Fig.~\ref{fig_nak41_aa_p} our model predicts
a series of $p$-wave Feshbach resonances at both weak and strong magnetic
fields in the absolute ground state. 
Tab.~\ref{tab_nak41_p} confirms
as expected that multiplet splittings at small $B$ are ``anomalous''
in the same sense as for boson-fermion pair, whereas like in
Na$^{39}$K they follow standard patterns at large $B$. 
\begin{table}[hb!t]
\caption[]{Same as Tab.~\ref{tab_nak39_p} but for Na$^{41}$K.}
\label{tab_nak41_p}
\begin{tabular}{p{7cm}ccp{7cm}}
\begin{ruledtabular}
\begin{tabular}{ccr}
Na$^{41}$K channel &$M$ & $B_\text{res}~(G)$  \\  \hline
$|1,1\rangle + |1,1\rangle$ & 1 & 17.82  	 \\
                            & 1 & 20.72  	 \\
                            & 1 & 31.47  	 \\
                            & 1 & 40.94  	 \\
                            & 1 & 133.49 	 \\
                            & 1 & 377.46	 \\
                            & 1 & 386.90 	 \\
                            & 1 & 418.20 	 \\
                            & 1 & 426.43 	 \\
                            & 1 & 438.89 	 \\
                            & 1 & 466.48 	 \\
                            & 1 & 478.98 	 \\
                            & 1 & 495.58 	 \\
\end{tabular}
\end{ruledtabular}
&~~~~~ & &
\begin{ruledtabular}
\begin{tabular}{ccr}
Na$^{41}$K channel & $M$ & $B_\text{res}~(G)$  \\ \hline
$|1,1\rangle + |1,1\rangle$ &2 & 17.79        \\
                            &2 & 20.47        \\
                            &2 & 41.89        \\
                            &2 & 377.84       \\
                            &2 & 417.19       \\
                            &2 & 427.37       \\
                            &2 & 453.54       \\
                            &2 & 465.11       \\
                            &2 & 480.30       \\ \hline
$|1,1\rangle + |1,1\rangle$ &3 & 17.84        \\
                            &3 & 418.19       \\
                            &3 & 452.11       \\
                            &3 & 466.47       \\
\end{tabular}
\end{ruledtabular}
\end{tabular}
\end{table}
Experimental observation of the corresponding magnetic spectra would provide a valuable
piece of information to confirm the accuracy of our model for
$\ell >0$ collision in this boson-boson mixture.

\clearpage
\section{Conclusions}\label{sec_conclusion}

We have presented an extensive compendium of the ground state scattering
properties of isotopic NaK mixtures in an external magnetic field. Our
results complement existing theory and experimental data on the
boson-fermion pair Na$^{40}$K. The Feshbach resonance locations
and strengths we predict for the boson-boson pairs should be of
major interest for experiments in which control of the atom-atom
interaction is a requirement. Our spin-resolved analysis of Feshbach
molecules also provides an important piece of information for designing
magnetoassociation and two-photon transfer scheme of Feshbach molecules
to the absolute ro-vibrational ground state.

\section*{Acknowledgments}
This work is supported by the Agence Nationale de la Recherche (Contract
COLORI No. ANR-12-BS04-0020-01).

\end{document}